\documentclass[10pt,journal,compsoc]{IEEEtran}
\usepackage{amsmath,amsfonts}
\usepackage{array}
\usepackage{subfig}
\usepackage{textcomp}
\usepackage{stfloats}
\usepackage{url}
\usepackage{verbatim}
\usepackage{graphicx}
\usepackage{cite}
\usepackage{multirow}

\newtheorem{definition}{Definition}
\usepackage[table,xcdraw]{xcolor}
\usepackage[ruled,vlined]{algorithm2e}

\begin{document}

\title{Update Selective Parameters: Federated Machine Unlearning Based on Model Explanation}

\author{Heng Xu,~
        Tianqing Zhu*,~\IEEEmembership{Member,~IEEE,} 
        Lefeng Zhang,~
        Wanlei~Zhou,~\IEEEmembership{Senior Member,~IEEE,}
        and~Philip~S.~Yu,~\IEEEmembership{Fellow,~IEEE}
        \thanks{*Tianqing Zhu is the corresponding author. Heng Xu is with the Centre for Cyber Security and Privacy and the School of Computer Science, University of Technology Sydney, Ultimo, NSW 2007, Australia (e-mail: heng.xu-2@student.uts.edu.au). Tianqing Zhu, Lefeng Zhang and Wanlei Zhou are with the City University of Macau, Macau (e-mail: tqzhu@cityu.edu.mo; lfzhang@cityu.edu.mo; wlzhou@cityu.edu.mo). Philip S. Yu is with University of Illinois at Chicago (e-mail: Psyu@uic.edu).}
        }



\maketitle

\begin{abstract}
Federated learning is a promising privacy-preserving paradigm for distributed machine learning. In this context, there is sometimes a need for a specialized process called machine unlearning, which is required when the effect of some specific training samples needs to be removed from a learning model due to privacy, security, usability, and/or legislative factors. However, problems arise when current centralized unlearning methods are applied to existing federated learning, in which the server aims to remove all information about a class from the global model. Centralized unlearning usually focuses on simple models or is premised on the ability to access all training data at a central node. However, training data cannot be accessed on the server under the federated learning paradigm, conflicting with the requirements of the centralized unlearning process. Additionally, there are high computation and communication costs associated with accessing clients' data, especially in scenarios involving numerous clients or complex global models. To address these concerns, we propose a more effective and efficient federated unlearning scheme based on the concept of model explanation. Model explanation involves understanding deep networks and individual channel importance, so that this understanding can be used to determine which model channels are critical for classes that need to be unlearned. We select the most influential channels within an already-trained model for the data that need to be unlearned and fine-tune only influential channels to remove the contribution made by those data. In this way, we can simultaneously avoid huge consumption costs and ensure that the unlearned model maintains good performance. Experiments with different training models on various datasets demonstrate the effectiveness of the proposed approach.
\end{abstract}


\section{Introduction}
\label{sec:introduction}

\IEEEPARstart{M}{achine} unlearning is an emerging technology that has come to attract widespread attention. As model training becomes increasingly dependent on data, a growing number of factors—including regulations and laws, along with privacy, security, and usability concerns—have resulted in the need to remove the impact of a sample or multiple samples from a trained machine learning model. This reverse process is referred to as \textit{machine unlearning}. As an illustrative example, regulations and laws such as GDPR~\cite{webpage:GDPR} grant users \textit{the right to be forgotten}; to achieve this goal, the unlearning process is necessary. Moreover, some machine learning attacks can reveal private information about specific contents of a training dataset~\cite{DBLP:conf/sp/ShokriSSS17}, which can seriously threaten users' privacy and lead to concerned users removing their data from dataset and its contributions from machine learning models. In addition, if a dataset contains a large amount of outdated or low-quality data, this may affect the model performance~\cite{DBLP:journals/tkde/ChanGHS22}, meaning that unlearning is required. Therefore, machine unlearning has become an urgent new requirement in various communities. 

The most direct approach to unlearning involves retraining the machine learning model from scratch by deleting the unlearning data in advance. However, as machine learning or deep learning typically now involves a huge amount of training data, the retraining method is too time-consuming and cost-ineffective~\cite{DBLP:conf/sp/CaoY15}. Recent studies have developed several types of unlearning methods, specifically data reorganization and parameter manipulation. Data reorganization refers to situations in which model providers simplify the unlearning process by reorganizing the training dataset in advance~\cite{DBLP:conf/aaai/GravesNG21,DBLP:conf/sp/BourtouleCCJTZL21}. For example, Bourtoule et al.~\cite{DBLP:conf/sp/BourtouleCCJTZL21} divide the training dataset into multiple subsets and train sub-models based on each of these subsets. When a group of samples needs to be unlearned, this can be done by retraining sub-models based on the corresponding subset that contains those unlearning data, rather than the entire training dataset. For parameter manipulation, model providers usually offset the effect of the samples that need to be unlearned by directly manipulating the model parameters~\cite{DBLP:conf/cvpr/GolatkarAS20, DBLP:conf/icml/GuoGHM20}. Both types of methods have been quite popular in recent years. 

In this work, we consider the problem of unlearning in the setting of federated learning, in which the server wants to remove all samples associated with a particular class. There are many scenarios where data from a trained model that needs to be unlearned belongs to one or more classes~\cite{DBLP:journals/corr/abs-2111-08947,DBLP:journals/ml/BaumhauerSZ22}. For example, if the training process of federated learning contains a group of outdated data or data identified as adversarial type, the server needs to remove all effects of these data from the global model. Another example is thats the recommendation system needs to periodically remove some items~(classes) due to relevant legal updates or product iterations. Considering this setting, the above-mentioned centralized unlearning methods are inefficient due to the following challenges.


For the data reorganization methods, most centralized unlearning algorithms make use of the original training dataset to remove the effect of unlearning data~\cite{DBLP:conf/sp/CaoY15,DBLP:conf/sp/BourtouleCCJTZL21,DBLP:conf/www/Chen0ZD22,DBLP:conf/ccs/Chen000H022}. Those kinds of unlearning methods usually aggregate all training data in one node and split those data into different subsets. Then, each subset will be used to train the sub-model, which limits the impact of data in the subset to the corresponding sub-models rather than a more complex model with the entire training data. In the federated learning context, however, the original training dataset may not be accessible on the server, meaning that the aforementioned approaches will be unsuitable for federated learning. 
For parameter manipulation-based schemes, those methods always focus on simple machine learning problems, such as tree-based models~\cite{DBLP:conf/sigmod/SchelterGD21,DBLP:conf/icml/BrophyL21}; notably, the complexity of global models and the stochastic nature of the local learning process make it impossible to use parameter manipulation-based methods in federated learning. Moreover, due to the complexity associated with calculating the impact of the specified sample, manipulating a model’s parameters based on unreliable impact results or assumptions will significantly affect the model's performance~\cite{DBLP:conf/cvpr/GolatkarAS20, DBLP:conf/icml/GuoGHM20}.

For existing class-level unlearning in a federated learning setting, Wang et al. introduced the concept of TF-IDF~\cite{DBLP:journals/asc/ThakkarC20} to quantify the class discrimination of the channels and accordingly proposed an unlearning scheme based on the model pruning technique. However, the pruning-based scheme not only removes the information of the unlearning classes but also erases the information of some remaining data at the same time, which will lead to the degradation of the model performance for the remaining data. Other unlearning schemes usually use the historical parameter updates cached during the global training process. However, those cached gradients will not be updated with the unlearning process, which precludes carrying out subsequent unlearning processes. This solution may also further increase the risk to users' data since recent studies show that attackers can reconstruct the training dataset by referring to the gradients~\cite{DBLP:conf/nips/HuangGSLA21}. Finally, compared with traditional centralized learning, federated learning has more randomness involved. For example, the selected users involved in each global iteration during the training process make the federated learning non-deterministic and difficult to control. Even if retraining the model from scratch causes the global model to converge to different local minima at different times. Therefore, it becomes impractical and more difficult to unlearn data by directly manipulating the parameters in a federated learning setting~\cite{DBLP:conf/infocom/LiuXYWL22}.

To tackle the above issues, we consider that the purpose of unlearning is to remove the effect of certain specific samples from the model, not all training datasets. Thus, it is crucial to find the partial parameters that most affect these samples. Model explanation refers to a series of methodologies that can help to explain the importance of particular parameters~\cite{DBLP:journals/tkde/LiCSBGQWGZXC22,DBLP:conf/cvpr/LiLZLDWHJ19}. It can be used to find some channels within a trained model that have the greatest effect on the performance of the samples that need to be unlearned. 

Based on this observation, we first calculate the influence of each channel based on the ablation study, then select some channels that most affect the classification performance of unlearning class. Subsequently, we propose two different unlearning schemes based on the knowledge owned by the server. The first is decentralized unlearning, which is based on the original federated learning paradigm and achieves the unlearning purpose by only updating partial parameters within selected important channels. When the server owns some samples with the same distribution of training data, we prefer the second unlearning scheme, centralized unlearning. In this case, we put the unlearning operation on the server side with the help of only a few training samples rather than the whole training dataset contained in all clients. Centralized unlearning also only fine-tunes partial parameters within a model to unlearn data. In addition, to speed up the unlearning process, we use perturbation samples in both schemes to efficiently muddy the model’s understanding of the unlearning data. Those perturbation samples are generated from unlearning data with randomly selected incorrect labels. Both schemes can avoid large communication and computation expenses and simultaneously maintain the performance of the unlearned model for the remaining data. 

The main contributions of this paper can be summarized as follows:




\begin{itemize}
    \item First, we introduce the ablation study to calculate the influence of each channel with respect to the unlearning class, which enables us to get the most influential channels for the unlearning class within any trained model.

    \item Second, we provide two different schemes based on the knowledge owned by the server, decentralized unlearning and centralized unlearning. Both schemes only update partial influential channels' parameters to avoid large communication, computation and storage expenses.

    \item Third, we propose an effective unlearning method based on perturbed samples, which can effectively accelerate the process of unlearning while ensuring the usability of the unlearned model.
    
    \item Further analysis and experiments comparing the performance of our proposed approach with that of recent works under different models and multiple datasets demonstrate the effectiveness of the proposed federated unlearning scheme.

\end{itemize}

\section{Related Work}
\label{sec:relatedwork}


\subsection{Machine Unlearning}

Machine unlearning has been a hot topic in recent years. In our previously published survey paper~\cite{10.1145/3603620}, we conducted a comprehensive survey encompassing recent studies on machine unlearning. The survey explored several critical aspects, including: (i) exploring the motivation behind unlearning, (ii) defining the objectives and desired outcomes of unlearning, (iii) creating a novel taxonomy for categorizing existing techniques based on their rationale and strategies, and (iv) approaches for verifying the effectiveness of machine unlearning. In general, most existing machine unlearning solutions can be broadly divided into one of two categories: data reorganization methods~\cite{DBLP:conf/sp/CaoY15,DBLP:conf/sp/BourtouleCCJTZL21,DBLP:conf/www/Chen0ZD22,DBLP:conf/ccs/Chen000H022} and parameter manipulation methods~\cite{DBLP:conf/cvpr/GolatkarAS20, DBLP:conf/icml/GuoGHM20, DBLP:conf/sigmod/SchelterGD21,DBLP:conf/icml/BrophyL21}.

Cao et al.~\cite{DBLP:conf/sp/CaoY15} were the first to consider the problem of removing the effect made by certain samples from an already trained model and coined the term \textit{machine unlearning}. Inspired by the previous work, MapReduce~\cite{DBLP:journals/tcbb/SinhaPD22}, Cao et al. transformed the learning algorithms into summation form and used these summations to construct statistical query (SQ) learning models. The unlearning operation recomputes summations containing samples that need to be unlearned to remove the effect of these samples. Bourtoule et al.~\cite{DBLP:conf/sp/BourtouleCCJTZL21} proposed a ``sharded, isolated, sliced, and aggregated'' ($\mathbf{SISA}$) framework, similar to the current distributed training strategies~\cite{DBLP:journals/tifs/GrattonVAW22}, as a method of machine unlearning. Chen et al.~\cite{DBLP:conf/www/Chen0ZD22} introduced the method in~\cite{DBLP:conf/sp/BourtouleCCJTZL21} to recommendation systems and designed three novel data partition algorithms. Those methods are used to divide the data into balanced groups in order to ensure that collaborative information is retained. The strategy established by Bourtoule et al.~\cite{DBLP:conf/sp/BourtouleCCJTZL21} also introduced to unlearn graph type data~\cite{DBLP:conf/ccs/Chen000H022}.

For parameter manipulation, Golatkar et al.~\cite{DBLP:conf/cvpr/GolatkarAS20} proposed using the fisher information~\cite{DBLP:journals/jmlr/Martens20} of the remaining dataset to unlearn specific samples, with noise injected to optimize the shifting effect.  Guo et al.~\cite{DBLP:conf/icml/GuoGHM20} proposed an unlearning scheme called certified removal, based on influence theory~\cite{DBLP:conf/icml/KohL17} and differential privacy~\cite{DBLP:journals/tkde/ZhuYWZY22}, which involves first computing the influence of the unlearning data on the trained model and then updating the model to remove that influence. Differential privacy is introduced to limit the maximum difference between the unlearned model and the retrained model. Schelter et al.~\cite{DBLP:conf/sigmod/SchelterGD21}, and Brophy et al.~\cite{DBLP:conf/icml/BrophyL21} focused on unlearning methods in tree-based models. These calculated almost all possible sub-models in advance during the training process, which were stored together with the deployed model. Subsequently, when an unlearning request arrives, only the sub-models affected by the unlearning operation need to be replaced with the pre-stored sub-models.

In the context of federated learning, Wang et al.~\cite{DBLP:conf/www/Wang0XQ22} proposed a class-level unlearning scheme via channel pruning~\cite{DBLP:conf/cvpr/GuoWLY20}, which is followed by a fine-tuning to recover the performance of the pruned model. Liu et al.~\cite{DBLP:conf/iwqos/LiuMYWL21} proposed a federated unlearning methodology, FedEraser, which is largely based on the idea of using storage on the central server to aid in constructing the unlearned model. Historical parameter updates from the clients are stored in the central server during the training process, after which the unlearning process is achieved via calibration training. Liu et al.~\cite{DBLP:conf/infocom/LiuXYWL22} transferred the unlearning method from a centralized environment to federated learning by proposing a distributed Newton-type model updating algorithm to approximate the loss function trained by the local optimizer on the remaining dataset. 

In summary, the reviewed unlearning works have two common limitations. First, most existing research on machine unlearning focuses on centralized learning scenarios. Those methods always have free access to training datasets~\cite{DBLP:conf/sp/CaoY15,DBLP:conf/sp/BourtouleCCJTZL21,DBLP:conf/www/Chen0ZD22,DBLP:conf/ccs/Chen000H022} or are not applicable to complex models~\cite{DBLP:conf/cvpr/GolatkarAS20, DBLP:conf/icml/GuoGHM20, DBLP:conf/sigmod/SchelterGD21,DBLP:conf/icml/BrophyL21}, such as deep neural networks (DNNs). Second, unlearning schemes in federated learning are usually based on historical parameter updates or approximated updates, which usually reduce the performance of the unlearned model~\cite{DBLP:conf/www/Wang0XQ22,DBLP:conf/iwqos/LiuMYWL21,DBLP:conf/infocom/LiuXYWL22}.

\subsection{Model Explanation}

The interpretability of neural networks is a concept that has received increasing attention in recent years~\cite{DBLP:journals/pami/ZhangWCWSZ21}. Different methods have been developed to explore the operating mechanism encoded inside a deep model. Morcos et al.~\cite{DBLP:conf/iclr/MorcosBRB18} found that when removed from the network, units that are important for one category do not seem to cause more damage than unimportant units. Based on this, Zhou et al.~\cite{DBLP:journals/corr/abs-1806-02891} further found that ablating individual units tends to cause significant damage to the generalization accuracy of a subset of classes. Furthermore, Meyes et al.~\cite{DBLP:journals/corr/abs-1901-08644} revealed the different contributions made to the overall classification performance through the ablation of individual units. These authors found that some units are generally important for the task, representing different features of most classes, while others are selectively important for specific classes. The findings further show that some units are not important at all, which provides the network with a certain degree of robustness to structural damage. Li et al.~\cite{DBLP:conf/cvpr/LiLZLDWHJ19} found that the importance of channels can vary even within the same layer. The more information the channel represents, the more important it is for the network.

\section{Problem Statement}
\label{sec:problemstatement}


\subsection{Federated Learning Background}
Federated learning is a decentralized machine learning framework,  the key components of which are a server $S$ and a group of clients $U = \left\{u_1, u_2,...,u_n\right\}$, where $n$ is the number of clients, as shown in Figure~\ref{fig:standardframework}. The server relies on these clients to collaboratively complete the training of a global model $M$, which will be provided to all clients to use after the training process. The model's training process will repeat the steps outlined below until the model satisfies certain conditions. At the $i$-th training round ($i \in E$, where $E$ is the number of global training rounds), the updating process of the global model $M$ is performed as follows:

\begin{figure}[!t]
\centering
\includegraphics[width=0.48\textwidth]{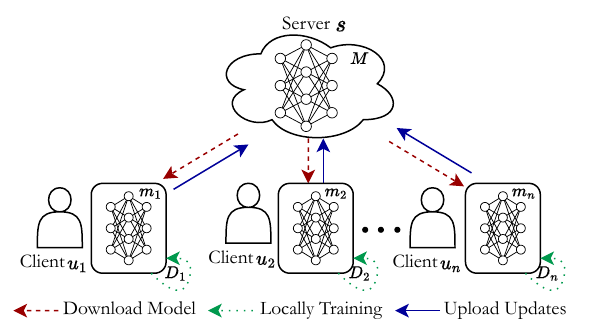}
\caption{A Standard Framework of Federated Learning.}
\label{fig:standardframework}
\end{figure}

\begin{enumerate}
    \item \textbf{Initialization:} The server first selects a certain percentage of clients to participate in this round of model training due to the unstable characteristics of clients in the federated learning framework. Subsequently, the selected federated clients download the current global model $M_i$ and the training configurations from the server.
    \item \textbf{Local Training and Uploading:} Each selected client $u_k$ trains the local model $m_k = M_i$, based on its own local dataset $D_k$ and training configurations for $E_{local}$ epochs, then computes an update with respect to $m_k$, $(k \in \left\{1, 2,...,n\right\})$, where $E_{local}$ is the number of local epochs for each client. When the local training process is complete, each client will upload its model updates to the server.
    \item \textbf{Server Aggregation:} Once the server has collected all updates from all participating clients, it will update the global model $M_i$ based on one aggregation rule, such as FedAvg~\cite{DBLP:journals/csur/LoLWPZ21}, and the collected updates to obtain a new global model $M_{i+1}$. $M_{i+1}$ will play the role of $M_i$ for the next training round.
\end{enumerate}

When the global model converges, the server halts the above training process and obtains the final federated learning model $M^{*}$.

\subsection{Examples: Federated Unlearning}
\label{subsection:examples}

Designing an effective and efficient unlearning scheme for a federated learning framework is difficult. First, the incremental learning process causes the global model updating to depend on all previous updates. If an update from one client is removed from the global model's aggregation process in round $i$, the global model will be changed after the aggregation process; all the client's subsequent updates will then become useless, as all clients compute local updates based on the global model. Second, randomness exists in federated learning training due to the stochastic training process; this impacts which clients will be participating in this round's training process, as well as what samples will be used for training in a certain epoch or batch. This randomness makes the federated learning process non-deterministic and difficult to control.

\begin{figure}[!t]
\centering
\includegraphics[width=0.45\textwidth]{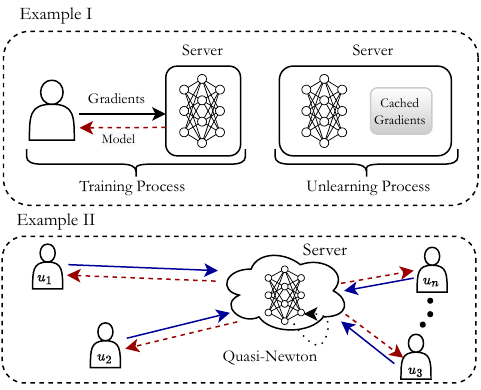}
\caption{Examples of Federated Unlearning.}
\label{fig:examples}
\end{figure}

To further illustrate the difficulties in federated learning, we present here two examples of existing unlearning schemes. Figure~\ref{fig:examples} illustrates two schemes commonly used in federated unlearning. In example I, the server utilizes cached historical gradients from clients in the training process to approximate the gradients in the unlearning process. Notably, the effectiveness of these schemes will decrease dramatically as the number of unlearning requests increases; this is because the gradients are cached during the training phase, and the unlearning process will not update these gradients to satisfy subsequent unlearning requests~\cite{DBLP:conf/iwqos/LiuMYWL21}. Second, the clients involved in each federated learning training process are randomly selected. The gradients cached during the training period may not be used in the unlearning process since the relevant clients may be different and the gradients are also not related to each other. Some recent studies have even shown that attackers can use the gradient of model updates to reconstruct clients' data; therefore, these solutions may increase the risk of data leakage from clients in federated learning~\cite{DBLP:journals/csur/YinZH21}. Example II introduces the parameter manipulation-based unlearning scheme from centralized learning to federated learning. Those methods usually use Quasi-Newton methods to update global models instead of the original model retraining process. Quasi-Newton technique is an optimization algorithm designed for unconstrained numerical optimization problems. It approximates the inverse Hessian matrix iteratively, making it suitable for large-scale optimization where direct computation of the Hessian is impractical. Nonetheless, similar to centralized learning, this method is only applicable to simple models with minimal data erasure.

The two examples above illustrate that it is impractical to achieve unlearning based on cached gradients or Quasi-Newton; if either of these schemes is used for unlearning data, it will reduce the performance of the unlearned model for the remaining data. 

\subsection{Federated Unlearning Formalization}
We consider federated unlearning in a supervised learning setting, i.e., one in which all clients collaboratively train a supervised model under the control of the server. Each client's data belongs to the instance space $\mathcal{X} \subseteq \mathbb{R}^{d}$, with the label space defined as $\mathcal{Y} \subseteq \mathbb{R}$. We consider the local data is Independent and Identically Distributed~(IID) and use $\mathcal{D}_{k}=\left\{\left(\mathbf{x}_{k,1}, y_{k,1}\right),...,\left(\mathbf{x}_{k,t}, y_{k,t}\right) \right\} \subseteq \mathbb{R}^{d} \times \mathbb{R}$ to represent a local training dataset in client $u_{k}$, in which each sample $\mathbf{x}_{k,t} \in \mathcal{X}$ is a $d$-dimensional vector, $y_{k,t} \in \mathcal{Y}$ is the corresponding label, and $t$ is the size of $\mathcal{D}_{k}$. To unlearn data, all clients first need to perform a \textit{data deletion} operation locally. 

\begin{definition}(\textit{Data Deletion}). Data deletion performed by each client is an operation that deletes a group of client data that needs to be unlearned from the global model, denoted as unlearning data $\mathcal{D}_{k}^{u}$. For each client $u_{k} \in U$, given its local dataset $\mathcal{D}_{k}$ and unlearning data $\mathcal{D}_{k}^{u}$, data deletion is defined as follows:

\begin{equation}
    \mathcal{D}_{k}^{r}= \mathcal{R}(\mathcal{D}_{k},\mathcal{D}_{k}^{u})
\end{equation}
\end{definition}

where the operation $\mathcal{R}$ denotes deleting data $\mathcal{D}_{k}^{u}$ from $\mathcal{D}_{k}$. For example, a client deletes data samples $\mathcal{D}_{k}^{u}$ from the local dataset $\mathcal{D}_{k}$ according to the designed data deletion operation in order to obtain the remaining dataset $\mathcal{D}_{k}^{r}$. The purpose of data deletion is to remove unlearning data from the training set in all clients so that the global model will not be affected by those data again in the next collaborative training process. After data deletion, the server will perform the \textit{federated unlearning} operations.

\begin{definition}(\textit{Federated unlearning}) Consider a set of data that we wish to remove from the global model, denoted as $\mathcal{D}_{u} = \left\{\mathcal{D}_{1}^{u}, \mathcal{D}_{2}^{u},...,\mathcal{D}_{n}^{u}\right\}$. Let $M^{*}$ denote the global model produced by the federated learning, while $M^u$ denotes the unlearned model that we want to achieve. We define \textit{federated unlearning} $\mathcal{U}\left( \cdot \right)$ as follows:
\begin{equation}
    M^u \equiv \mathcal{U}(M^{*}, \mathcal{D}_{u} )
\end{equation}
\end{definition}

Federated unlearning aims to remove the contribution about the unlearning data from the global model. After the unlearning process is complete, a verification function $\mathcal{V}(\cdot)$ can be used to measure whether model $M^{u}$ has successfully unlearned the requested unlearning data.

\begin{definition}(\textit{Verification function})
    After the unlearning process is complete, a verification function $\mathcal{V}(\cdot)$ can perform a distinguishable check—that is, $\mathcal{V}\left(M^u\right) \neq \mathcal{V}\left(M^{*}\right)$—to illustrate the results of the unlearning operation.
\end{definition}

As an attack method, membership inference attacks can also be used to determine whether a sample's contribution has been successfully removed from the model~\cite{DBLP:conf/sp/ShokriSSS17}. 

\begin{table}
  \caption{Notations}
  \renewcommand{\arraystretch}{1.2}
  \label{tab:notations}
  \centering
  \begin{tabular}{c|c}
    \hline
    Notations &  Explanation \\
    \hline
    $u_k$                                       &One client\\
    $D_k$                                       &The dataset in $u_k$	\\
    $\left( \mathbf{x}_{k,t}, y_{k,t} \right)$  &One sample in $D_k$\\
    $\mathcal{R}$                               &The data deletion operation\\
    $D_k^{u}$                                   &The unlearning dataset in $u_k$	\\
    $D_k^{r}$                                   &The remaining dataset in $u_k$	\\
    $\mathcal{U}$                               &The federated unlearning process\\
    $M^{*}$                                     &The learned federated learning model\\
    $M^u$                                       &The unlearned model\\
    $\mathcal{V}$                               &The unlearning operation\\
    $y^{u}$                                     &The unlearning class label\\
    $S_{w_{l,i}}$                               &The effect of channel $w_{l,i}$ in layer $l$\\
    $H(M, D)$                                   &The model accuracy of dataset $D$ \\
    $\mathcal{T}$                               &The important channels\\
    \hline
    \end{tabular}
\end{table}

\subsection{Design Goals}

We provide a federated unlearning scheme based on the concept of model explanation, which can achieve federated unlearning quickly while effectively ensuring the performance of the unlearned global model. The desired goals of our proposed unlearning scheme are as follows:

\begin{itemize}
    \item \textbf{Effective}: Similar to retraining from scratch on the remaining data, our scheme should ensure that the unlearned model does not contain any information about the unlearning data (i.e., that attackers cannot infer information about unlearning data through the unlearned model using existing privacy attack methods, such as membership inference attack~\cite{DBLP:conf/sp/ShokriSSS17}).

    \item \textbf{Efficient}: Regardless of how much data needs to be unlearned, our scheme should be more efficient than retraining from scratch; in short, the unlearning solution should maintain communication, computation and storage costs at reasonable levels.     
    
    \item \textbf{Accuracy}: Even though unlearning operations will cause a reduction in model performance, our scheme should not unduly reduce performance through unreasonable manipulation.
    
    \item \textbf{Security}: It should not introduce new privacy risks and at least be consistent with the original guarantee. Specifically, it should ensure that any attacker, including the server, cannot easily utilize gradient information to implement gradient leakage attacks.
    
    \item \textbf{Non-intrusive}: Our scheme should be non-intrusive for both server and clients, and should be able to serve as an opt-in component inside existing federated learning systems; that is to say, the operation of unlearning cannot affect the current structure of the federated learning system. 
\end{itemize}

It is worth noting that security goal emphasizes that the designed unlearning scheme cannot introduce potential threats that additionally threaten the user's privacy, while effective goal underscores that no one can recover unlearned data from the model after unlearning process.

\section{Federated Unlearning Methodology}
\label{sec:methodology}
\subsection{Overview}
In this section, we first present an overview of our proposed scheme, then provide a detailed outline. For ease of reference, the important notations that appear in this paper and their corresponding descriptions are listed in TABLE~\ref{tab:notations}.

\begin{figure}[!t]
\centering
\includegraphics[width=0.46\textwidth]{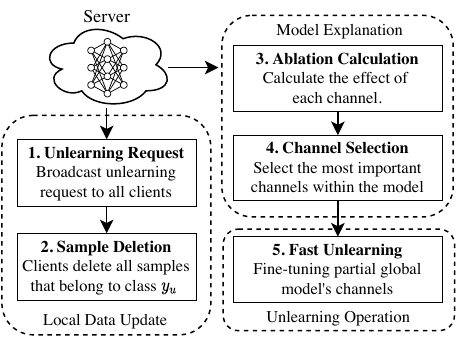}
\caption{Overview and Workflow of Our Federated Unlearning Scheme.}
\label{fig:scheme}
\end{figure}

In this paper, we focus on unlearning requests relating to a specific class—that is, requests designed to remove an entire class of samples from the global model, which is frequently required in unlearning problems~\cite{DBLP:conf/aaai/GravesNG21,DBLP:conf/www/Wang0XQ22}. Figure~\ref{fig:scheme} illustrates the workflow of the proposed unlearning method, which mainly comprises five steps. We will discuss the first two steps in the local data update section, after which ablation calculation and channel selection will be discussed in the model explanation section. The unlearning schemes are presented in the unlearning process section, which contains two different sub-schemes for different application settings.

Our key insight is that each channel of a model contributes differently to the overall classification performance. This means that some channels are universally important for the task, while others are selectively important for specific classes. Based on this observation, we use the ablation study to find the most influential channels that have the greatest effect on the unlearning class. After that, we fine-tune those influential channels to remove the contribution of unlearning data from the global model. Since the number of parameters in influential channels is much less compared to the number of all model parameters, the unlearning operation could avoid extensive communication between clients and a server. At the same time, because fewer parameters need to be updated, the unlearning operation will be more efficient than retraining all parameters. We also establish a benchmark to control how many influential channels need to be fine-tuned. Broadly speaking, the fewer channels are fine-tuned, the more efficient unlearning occurs.

\subsection{Local Data Update}
We assume that in the initialization stage of the federated learning, the server aims to train a deep neural network model with $L$ layers. The parameters in each layer can be represented as $\mathbf{w}=\left\{w_{1}, \ldots, w_{L}\right\}$. If layer $l$ is a convolutional layer, the parameters $w_{l}$ in the $l$-th layer is denoted as $\mathcal{R}^{C_{l}^{out} \times C_{l}^{in} \times K_{l} \times K_{l}}$, where $C_{l}^{out}, C_{l}^{in}$ and $K_{l}$ denote the number of output channels, number of input channels, and the kernel size, respectively. If layer $l$ is a linear layer, the parameters $w_{l}$ can be represented as $\mathcal{R}^{C_{l}^{out} \times C_{l}^{in}}$.

During the model training or at the inference stage, the server can broadcast unlearning requests to all clients. First, when the server gets an unlearning request, it will generate the unlearning requests $(u_k, y^{u}), u_k \in U$ and send those pairs to each corresponding client. After receiving the unlearning request from the server, all clients should remove the samples that need to be unlearned from $D_k$ to prevent these from being re-included in subsequent stages of the training process. That is, $D_{k}^{r} = \mathcal{R}(D_{k}, \left( \mathbf{x}_i, y^{u} \right))$, for $\forall \left(\mathbf{x}_i, y^{u}\right) \in D_{k}$, where $y^{u}$ denotes the class label to be unlearned, and $\mathbf{x}_i$ represents a sample that belongs to $y^{u}$. 

\begin{figure}[!t]
\centering
\includegraphics[width=0.48\textwidth]{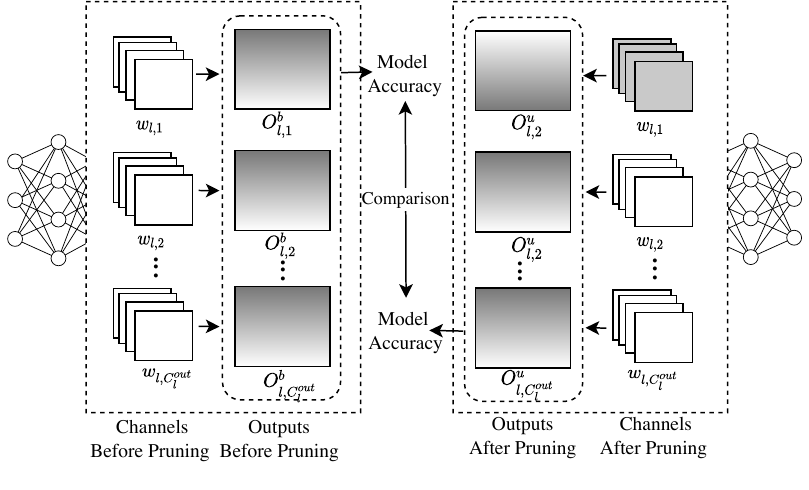}
\caption{Ablation Study.}
\label{fig:importance}
\end{figure}

\subsection{Model Explanation}

Machine learning models usually consist of multiple different units, and each of those units provides various contributions to the model's performance. Different from previous methods~\cite{DBLP:conf/iclr/MorcosBRB18,DBLP:journals/pami/ZhangWCWSZ21}, which usually use the ablation study to measure the contributions of those units to the overall performance, we introduce ablation study to analyze the channel influence for the performance of unlearning class and define this influence as the \textit{Effect of Channel}.

\begin{definition}(\textit{Effect of Channel})
    \label{def:effectofindividual}
    Given a global model $M^{*}$ that takes a group of data $I$ and outputs an overall accuracy $H(M^{*}, I)$. For a known baseline $H_{b}(M^{*}, I)$ that achieved before pruning operation, the influence $S_{w_{l,i}}$ of one channel $w_{l,i}$ in layer $l$ toward data $I$ is the change of the accuracy after pruning parameter in $w_{l,i}$ to obtain $M^{'}$:
    \begin{equation}
        S_{w_{l,i}} =  H_{b}(M^{*}, I) - H(M^{'}, I)
    \end{equation}
\end{definition}

Figure~\ref{fig:importance} illustrates one example of the calculation of the effect of a single channel within one convolutional layer $l$. We use $I$ to denote the input of the model and $I_{l}$ to represent the input of the $l$-th layer. Given the parameters $w_{l}$ in layer $l$, the output of the $l$-th layer in an original trained model can be represented as $O_{l}^{b}=I_{l} \otimes w_{l}$, where $\otimes$ is the convolutional operation. When pruning one specific channel $w_{l,1}$ in layer $l$, $w_{l}^{'} = \mathcal{P}(w_{l}, w_{l,1})$, where $\mathcal{P}$ is the pruning operation, the output will be changed to $O_{l}^{u} = I_{l} \otimes w_{l}^{'}$ and $O_{l}^{b} \neq O_{l}^{u}$. This difference of the single layer will be passed forward and will be reflected in the final accuracy $H(M^{'}, I)$. We calculate the effect according to the corresponding pruned channel based on the change of the accuracy $S_{w_{l,i}} = H_{b}(M^{*}, I) -  H(M^{'}, I)$. After achieving the effect of all channels, the server then selects the channels within a model that most affect the performance of the unlearning class, i.e., the channels with high effect values. We select these channels and consider them to be the most influential channels for an unlearning class. 

\begin{algorithm}[!t]
	\small
	\caption{Influential Channel Selection}
        \label{algorithm:ablationanalysis}
	\LinesNumbered 
	\KwIn{global model $M^{*}$, selection factor: $\delta$}
	\KwOut{important channels in model $M^{*}$: $\mathcal{T}$}
	~~~~~Initialize effect of channel $\mathcal{S}$ as an empty $\emptyset$\\
        ~~~~~Initialize $\mathcal{T}$ as an empty directory $\emptyset$\\
	~~~~~Construct a group of unlearning data $I$\\
	~~~~~Feed $I$ to model $M^{*}$ and record the original model accuracy $H_{b}(M^{*}, I)$\\
	~~~~~$\mathcal{E} \leftarrow $ all \textit{layers} in model $M^{*}$\\
	~~~~~\For{\rm each layer $e$ in $\mathcal{E}$}{
	    ~~~~~$w_{l} \leftarrow $ all \textit{channels} in $e$\\
        ~~~~~\For{\rm each channel $w_{l,i}$ in $w_{l}$}{
	        ~~~~Prune the channel $w_{l,i}$ to get $M^{'}$\\
                ~~~~~Feed $I$ and record accuracy $H(M^{'}, I)$\\
	        ~~~~~$S_{e_{l,i}} = H_{b}(M^{*}, I) - H(M^{'}, I)$\\
             }  
            ~~~~~~Sort $S_{e_{l}}$ in the order of $S_{e_{l,i}}$ from small to large\\
            ~~~~~~$\mathcal{T}_{e}$ $\leftarrow$ select the top $\delta$ channels\\
	}
	\Return {$\mathcal{T}$}\\
\end{algorithm}

Algorithm~\ref{algorithm:ablationanalysis} provides the selection of important channels within a global model. In the first step (Lines 3-4), the server feeds a group of unlearning data and records the original model accuracy. Then, in lines 5–11, the server calculates the effect for all channels. Line 6 indexes each layer in the global model, while line 8 indexes each channel in one layer. After pruning a particular channel, the model will recalculate the accuracy of $I$ to get the effect of this channel~(lines 9-11). A high effect value indicates that the corresponding channel plays an influential role in model performance with respect to $I$, while a low effect value means that this channel may be useless for $I$ only or for all training data. In lines 12-13, the server selects the most influential channels based on the effect value $S_{e_{l}}$ in layer $l$. $\delta$ is a hyperparameter used to control the proportion of selected channels. It is important to note that the above steps can be performed at any time during the inference phase; thus, the time consumption of the unlearning process is not increased.

\subsection{Unlearning Process}
Considering different application scenarios, we provide two schemes to fine-tune the global model. The first is to use the traditional federated learning architecture to complete the unlearning process when clients and all training data are available. When clients and training data are unstable, we employ server-based unlearning without any communicating with clients. Both schemes only update partial channels.

\begin{figure}[!t]
\centering
\includegraphics[width=0.48\textwidth]{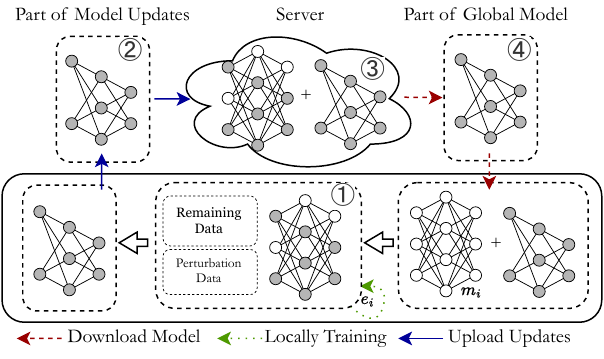}
\caption{Decentralized Unlearning.}
\label{fig:decentralizedunlearning}
\end{figure}

\subsubsection{Decentralized Unlearning}
If clients and those training data are available, the server will perform decentralized unlearning processes after selecting the important channels. As shown in Figure~\ref{fig:decentralizedunlearning}, there are three main differences between decentralized unlearning and federated learning. First, for the local training process, each client will only train the most influential channels based on two types of data~(\textcircled{1} in Figure~\ref{fig:decentralizedunlearning}). One is the remaining data, and the other is perturbation data, generated from unlearning data by modifying labels to an arbitrary value. Perturbation data can quickly obfuscate the model's understanding of the unlearning data, which will speed up the process of unlearning process. Second, in decentralized unlearning, each client only uploads influential channels' updates to the server for aggregation~(\textcircled{2} in Figure~\ref{fig:decentralizedunlearning}), while it will upload all channels' updates in federated learning. Third, in each step of aggregation, the server only updates partial channels based on the partial updates from clients and broadcasts the partial global model to each client for subsequent model training and updates~(\textcircled{3} and \textcircled{4} in Figure~\ref{fig:decentralizedunlearning}). In federated learning, the server updates all channels and broadcasts the whole model to each client. It is worth noting that even with the partial upload of the updates, the pattern of our unlearning scheme is the same as that of federated learning. The unlearning process can be done exactly based on the original federated learning paradigm.

Algorithm~\ref{algorithm:decentralizedunlearning} illustrates the process of decentralized unlearning. First, the server initializes the global model $M_{u}^{0}$ with the original global model $M^{*}$ and broadcasts part of the global model to all clients~(lines 3-4). Then, for each round of the unlearning process, each client downloads and updates the local model~(line 10). During each model training step, the unimportant channels $M \backslash \mathcal{T}$ are fixed, and only the important channels $\mathcal{T}$ are trained based on two types of data, remaining and perturbation data~(lines 11-14). After the local training process of a client, the client will upload only part of the model updates to the server~(line 15). When the server receives all partial updates from all clients, it will aggregate them and then broadcast a new partial global model to all clients for the next training~(lines 6-8). When the global model converges, $M^{u}$ is the model that has erased the information of the unlearning data.

\begin{algorithm}[!t]
	\small
 	\caption{Decentralized Unlearning}
	\label{algorithm:decentralizedunlearning}
	\LinesNumbered 
	\KwIn{important channels $\mathcal{T}$, original global model $M^{*}$}
	\KwOut{unlearned model $M^{u}$}
        Server broadcast important channels $\mathcal{T}$ to all clients\\
        \textbf{Server executes:}\\
        ~~~~~initialize $M_{u}^{0} \leftarrow M^{*}$\\  
        ~~~~~broadcast part of global model $M_{u,p}^{0}$ to all clients\\  
	~~~~~\For{\rm  $t$ from $1$ to $E_{global}$}{
            ~~~~~\For{\rm each client $u_{k}$ in $U$}{
                    $\mathcal{R}^{k} \leftarrow Clientupdate(k,M_{u,p}^{t})$\\
                    }
            ~~~~~$M_{u,p}^{t+1} \leftarrow M_{u,p}^{t} + \sum_{k=1}^{n}\frac{1}{n}\mathcal{R}^{k}$
	    }
        \textbf{Clientupdate(}$k,M_{u,p}^{t}$\textbf{):}  //Run one client k\\
        ~~~~~Update the local model $M^{t}$ based on $M_{u,p}^{t}$\\
  	~~~~~\For{\rm  each local epoch $e$ from $1$ to $E_{local}$}{
            ~~~~~\For{\rm each batch $b$ in $D_{k}$}{
                ~~~~~fix the channels in $M^{t} \backslash \mathcal{T}$\\
                ~~~~~only update channels $\mathcal{T}$ based on two types of data\\
	       }
            }
        upload the updates of $\mathcal{R}^{k}$ of channels $\mathcal{T}$ to server\\
\end{algorithm}

\subsubsection{Centralized Unlearning}
A fewer number of model channels to update usually means less training data is required. Therefore, we also provide a scheme that requires a small amount of training data on the server side to achieve the unlearning purpose, centralized unlearning. Our centralized unlearning process is similar to the original training procedure for each federated learning client. In particular, we fix non-influential channels within the model $M^{*} \backslash \mathcal{T}$. Then, in each model training process, we only update the important channels $\mathcal{T}$ based on remaining and perturbation data to reduce the computational cost. In addition, the process is only executed on the server without needing any client participation, which reduces communication and computation costs. 

\subsection{Case Study}

To further illustrate our unlearning method, we present a case study in this section. We assume that there are ten clients $U=\left\{u_{1}, u_{2}, \ldots, u_{10}\right\}$ training a federated learning model, ResNet20, each of which has certain data from dataset CIFAR-10, and denoted as $\mathcal{D}_{k}$. We assume that during the model training or inference stage, the server needs to remove all contributions related to class $1$ from the model. In unlearning request stage, the server generates unlearning request $(u_k, 1), u_{k} \in U$ and sends it to the corresponding client. Each client will remove all samples from the local training data after receiving the unlearning request.

At the same time, the server first selects the channels of the current model that most affect the classification performance of class $1$. To do this, the server constructs a small batch of data $I$ consisting of samples from class $1$, and feeds it into the model, recording original model accuracy $H_{b}(M^{*}, I)$. Then, given one layer $l$, the server prunes one channel $w_{l, i}$ at a time and retests the accuracy $H(M^{'}, I)$ of the above data $I$. Based on two accuracy results, the server will calculate the effective value of $w_{l, i}$, $ S_{w_{l,i}} = H_{b}(M, I) - H(M^{'}, I)$. This process is executed consistently until all channels in layer $l$ are indexed.  After that, the top $\delta$ channels with the higher effective value are selected; these will be denoted as the most influential channels in this layer. The same process is then performed for the other layers. 

After the above steps, the server will find the channels within a model that exert the greatest influence over the unlearning class $1$. The server will then select the unlearning process based on the availability of clients and those data. If the clients and those training data are available, the server will execute the decentralized unlearning process, where the local model update of each client as well as the uploaded update of each client only consider the most influential channels. If those clients or those data are unavailable, the server will execute the centralized unlearning process. In each step, the server holds some channels as fixed and only updates the influential channels to complete the unlearning. Our unlearning scheme is mainly based on the use of model explanation to select and fine-tune the most influential channels rather than approximating the unlearned model based on the gradients cached during the training phase. As a result, we can effectively avoid the problems discussed in Section~\ref{subsection:examples}. In addition, our schemes only update partial channels with the help of the perturbation data,  which can significantly accelerate the unlearning process.

In addition, our solution does not introduce any security risks and is non-intrusive. Unlearning requests can be sent through encrypted channels; even if such a request is obtained by a malicious attacker, this will not cause serious privacy concerns since the information contained in the request is less than that contained in the gradient. Furthermore, our method only uploads partial model channels during the decentralized unlearning or does not upload any gradient information in the centralized unlearning process, which can reduce the security risks introduced by gradients. Moreover, our scheme is non-intrusive and does not affect the traditional federated learning structure. To further improve the security of the training process, our scheme is also compatible with other schemes, such as differential privacy~\cite{DBLP:journals/tkde/ZhuYWZY22}. Therefore, our scheme does not decrease the security of traditional federated learning, or is at least consistent with the security of traditional federated learning, and can serve as an opt-in component within existing systems.

\section{Performance Analysis}

In this section, we analyze the performance of the proposed unlearning scheme in terms of computational, communication, and storage overheads. We term the decentralized unlearning as $\mathrm{Our}_{DE}$ and denote the centralized unlearning scheme as $\mathrm{Our}_{CE}$. Table~\ref{tab:techniquecomparison} shows the comparison results.

\begin{table*}[!htb]
    \centering
    \renewcommand{\arraystretch}{1.3}
    \caption{Comparison Results Between Our Scheme and Retraining From Scratch ($\delta < 1$)}
    \label{tab:techniquecomparison}
    \begin{tabular}{cccc}
    \hline
     \rowcolor{gray!20}                       & Computation Overhead & Communication Overhead & Storage Overhead     \\ \hline
      Retraining From Scratch & $n * (n^3 * 6f(\cdot) + g(\cdot))$                                        & $2 * n^2 * c(\cdot)$                & $\frac{|Y|- 1}{|Y|}*s(\cdot)$  \\
     \rowcolor{gray!20}                   $\mathrm{Our}_{DE}$                          & $n * (n^3 * f(\cdot) * (1 + 5 * \delta) + g(\cdot))$ & $2 * n^2 * c(\cdot) * \delta$       & $\frac{|Y|- 1}{|Y|}*s(\cdot)$ \\
      $\mathrm{Our}_{CE}$                          & $n^2 * f(\cdot) * (1 + 5 * \delta)$                  & $0$                                 & $0.05*s(\cdot)$  \\ \hline
    \end{tabular}
\end{table*}
\subsection{Computational Overhead}

The model training process is the most time-consuming element of a federated learning system. In the interests of simplicity, we set $f(\cdot)$ to represent the computational cost of the forward propagation, then the computational cost of one-step backpropagation is at most $5f(\cdot)$~\cite{DBLP:conf/infocom/LiuXYWL22}. Since the server also uses one aggregation algorithm during the training process to aggregate updates and update the global model, we use $g(\cdot)$ to denote the computational cost of the aggregation operation. 

For retraining from scratch, during the forward and backpropagation operation in the model updating for one batch, all channels will be involved, which costs $6f(\cdot)$. Moreover, it requires a group of users to participate in the unlearning process; this one batch model update should be assigned to all users and requires an additional aggregation operation. Therefore, the total computational cost of this scheme is $E_{global} * (\mathcal{N}_{user} * E_{local} * \mathcal{N}_{batch} * (6f(\cdot)) + g(\cdot))$. For the sake of convenience, we rewrite it as  $n * (n^3 * 6f(\cdot) + g(\cdot))$. For $\mathrm{Our}_{CE}$, during the forward propagation operation in one batch model update, all model channels are also involved in the calculation until the predicted value is output in the last layer, which costs $f(\cdot)$ units of computation. In the backpropagation operation, since we only focus on updating part of the channels, the cost will be $\delta * 5f(\cdot)$ units of computation, where $\delta$ is the proportion of important channels and $\delta < 1$. In addition, since neither requires client-server interaction in the unlearning process, there is no computation consumed by the aggregation operations. Finally, the total computational cost of our unlearning scheme $\mathrm{Our}_{CE}$ is $n^2 * f(\cdot) * (1 + 5 * \delta)$.  For our scheme $\mathrm{Our}_{DE}$, since we also only consider the update of important channels, so the total computational cost is $n * (n^3 * f(\cdot) * (1 + 5 * \delta) + g(\cdot))$.

Therefore, compared with retraining from scratch, our decentralized unlearning scheme $\mathrm{Our}_{DE}$, can achieve $\frac{n * (n^3 * 6f(\cdot) + g(\cdot))}{n * (n^3 * f(\cdot) * (1 + 5 * \delta) + g(\cdot))} \times$ unlearning improvement. For centralized unlearning scheme $\mathrm{Our}_{CE}$,  this ratio can be up to $\frac{n * (n^3 * 6f(\cdot) + g(\cdot))}{n^2 * f(\cdot) * ( 1+ 5 * \delta)} \times$. Assuming that the computation cost for aggregation operation $g(\cdot)$ is negligible and all clients calculate one batch model update in a parallel paradigm, both schemes can provide a speed-up of $\frac{6}{1 + 5 * \delta} \times$ and $\frac{6 * n^2}{(1 + 5 * \delta)} \times$, respectively. In addition, since we keep the information from previous training rather than retraining from scratch and using the perturbation data to accelerate the unlearning process, we can further improve our computational efficiency, e.g., the global epoch will be less than retraining from scratch. In summary, it is shown that our unlearning scheme significantly reduces the computational overheads compared with retraining from scratch.

\subsection{Communication Overhead}

There are normally two entities in an unlearning scheme, namely, the clients and the server. Without loss of generality, we evaluate communication efficiency through the number of client-server interactions. Specifically, we set $c(\cdot)$ to represent one-time interaction between a client and server.   

For retraining from scratch, it is necessary to interact with clients to complete the unlearning process. During each unlearning iteration, the server first broadcasts the current global unlearning model to each client, after which all clients will upload the unlearning information to the server; thus, retraining from scratch will spend $E * \mathcal{N} * \left( 2 * c(\cdot) \right)$, where $E$ and $\mathcal{N}$ are the number of unlearning global epochs and clients that participate in the unlearning process, respectively. For convenience, we also simplify this to $2 * n^2 * c(\cdot)$. For our decentralized unlearning scheme $\mathrm{Our}_{DE}$, since each client only sends updates about important channels, it will cost $2 * n^2 * c(\cdot) * \delta$. For centralized unlearning scheme $\mathrm{Our}_{CE}$, it places most of the computation on the server side, which precludes the need for interaction; thus, it almost costs $0$.

Therefore, compared with retraining from scratch, our decentralized unlearning scheme can save $\frac{1}{\delta} \times$ in terms of communication costs. For $\mathrm{Our}_{CE}$, since it does not communicate with the server, it has almost no consumption due to communication. Hence, our unlearning scheme is more efficient than retraining from scratch.

\subsection{Storage Overhead}

Unlearning methods are usually based on the participation of the remaining data. We consider the storage cost based on the size of the remaining data. Specifically, we use $s(\cdot)$ to denote the size of the original training data and use the $|Y|$ to represent the number of classes in training data.

For retraining from scratch and our decentralized unlearning scheme $\mathrm{Our}_{DE}$, since they need all remaining data of all clients to finish the unlearning process, it will cost $\frac{|Y|- 1}{|Y|}*s(\cdot)$ training samples. For our centralized unlearning scheme $\mathrm{Our}_{CE}$, since we put the unlearning process on the server side and only update partial channels' parameters, it only needs a small group of the remaining data. In our experimental results, for the CIFAR-10 dataset, when unlearning class $1$ from the model ResNet20, it only needs less than $0.05 * s(\cdot)$ size of original data. Therefore, compared with retraining from scratch, our scheme $\mathrm{Our}_{CE}$ largely reduces storage overheads.

\textit{Summary}: From the above analysis, the proposed unlearning schemes are determined to be efficient in terms of computational, communication, and storage costs compared with retraining from scratch, which demonstrates the practical potential and significant performance improvements obtained by our unlearning schemes. 

\section{Experiments}
\label{sec:expriments}


\subsection{Experimental Setup}


\subsubsection{Dataset and Model}

During the federated learning initialization phase, the server needs to broadcast the model to each client. Here, we choose the two most powerful and popular existing models to evaluate our unlearning, namely VGG~\cite{DBLP:journals/corr/SimonyanZ14a} and ResNet~\cite{DBLP:conf/cvpr/HeZRS16}. We also adopt three real-world image datasets for evaluation: MNIST~\footnote{http://yann.lecun.com/exdb/mnist/}, Fashion MNIST~\footnote{http://fashion-mnist.s3-website.eu-central-1.amazonaws.com/}, CIFAR-10 and CIFAR-100~\footnote{https://www.cs.toronto.edu/~kriz/cifar.html}. The datasets cover different attributes, dimensions, and numbers of categories, allowing us to explore the unlearning utility of the proposed algorithm effectively.

\subsubsection{Evaluation Metrics}
Here, we introduce the following criteria to evaluate the efficiency and effectiveness of our unlearning scheme:

\begin{itemize}
    \item \textbf{Training Rounds}: The number of training rounds indicates whether the model can complete the unlearning as quickly as possible. On the premise that the same unlearning effect is achieved, fewer training rounds can show that the scheme is more efficient.
 
    \item \textbf{Accuracy-based}: Models that have been trained usually have high prediction accuracy for trained samples, which means that the unlearning process can be verified by the accuracy of the model output. Ideally, for data that needs to be unlearned, the accuracy should be the same as a model trained without seeing the unlearning data. This could be infinitely close to zero. In addition, for the remaining data, the accuracy should almost keep unchanged.
    
    \item \textbf{Attack-based}: The basic purpose of unlearning is to reduce the risk of sensitive information leakage. Therefore, certain attack methods can be used to directly and effectively verify the success of unlearning operations. Here, we use model inversion attack~~\cite{DBLP:conf/ccs/FredriksonJR15} and membership inference attacks~(MIAs)\cite{DBLP:conf/csfw/YeomGFJ18} to evaluate our scheme. Ideally, before the unlearning, model inversion attack can effectively recover the unlearned data, while after the unlearning, it cannot recover the unlearned data. For MIAs, we use the success rate of attacking unlearning samples as our metric, that is, $\text{recall}= \frac{TP}{TP + FN}$, where $TP$ denotes the number of unlearning samples predicted to be in the training set and $TP + FN$ represents the total unlearning samples. Ideally, $\text{recall}$ on these unlearning samples should be close to 100\% before unlearning. After unlearning, $\text{recall}$ should be close to $0\%$. 

    \item \textbf{GradAttack}: In order to evaluate whether updating partial channel parameters can mitigate gradient-related risks. We use GradAttack~\cite{DBLP:conf/nips/HuangGSLA21} to evaluate the training process of retraining from scratch and our decentralized unlearning, respectively. Ideally, during the retraining from scratch, GradAttack can effectively recover relevant information based on gradients, whereas for our unlearning scheme, it should be unable to recover that information.
\end{itemize}

It is worth noting that the purpose of using GradAttack differs from above attack-based metrics. It is not used to demonstrate the effectiveness of our unlearning scheme but, rather, to illustrate that our partial parameter updating can, to some extent, mitigate attacks based on gradient information, which was not considered in exiting unlearning solutions.

\begin{table*}[!t]
\centering
\renewcommand{\arraystretch}{1.3}
\caption{The Model Accuracy after Pruning Different Types of Parameters.}
\label{tab:accuracyforlayer}
\begin{tabular}{ccccccc}
\hline
\rowcolor{gray!20}        & \multicolumn{2}{c}{\textit{block\_2.conv\_bn2.conv} in ResNet20} & \multicolumn{2}{c}{\textit{block\_2.conv\_bn1.conv}  in ResNet20} & \multicolumn{2}{c}{\textit{feature.conv\_7}  in VGG11} \\ \hline
                                            & Remaining    & Unlearning   & Remaining    & Unlearning    & Remaining     & Unlearning    \\
\rowcolor{gray!20}          Original Model  & 80.48\%           & 87.10\%           & 80.48\%           & 87.10\%            & 75.90\%             & 94.60\%            \\
                            Random          & 69.48\%           & 74.60\%           & 67.30\%           & 83.40\%            & 72.95\%            & 79.90\%            \\
\rowcolor{gray!20}          Non-Important  & 68.34\%           & 89.45\%           & 46.37\%           & 85.56\%            & 62.02\%            & 93.20\%            \\
                            Important      & 70.62\%           & 38.50\%           & 65.14\%           & 2.40\%             & 72.21\%            & 7.30\%            \\ \hline
\end{tabular}
\end{table*}

\subsubsection{Comparison Methods}
In our experiments, we respectively evaluate our decentralized unlearning~(\textbf{DE Importance}) and centralized unlearning~(\textbf{CE Importance}) schemes. To demonstrate the effectiveness of selecting the important channels, we also evaluate the validity of randomly selected channels in a decentralized unlearning setting~(\textbf{DE Random}). We compare our unlearning schemes with four unlearning methods applied to a federated learning system: retraining from scratch (\textbf{Fully Retraining}), pruning-based (\textbf{Pruning})~\cite{DBLP:conf/www/Wang0XQ22}, Federaser~\cite{DBLP:conf/iwqos/LiuMYWL21} and calculations-based (\textbf{The Right})~\cite{DBLP:conf/infocom/LiuXYWL22}). In~\cite{DBLP:conf/www/Wang0XQ22}, Wang et al. pruned the most important parameters and then fine-tuned the pruned model to recover the model performance. Liu et al. proposed an unlearning scheme in~\cite{DBLP:conf/iwqos/LiuMYWL21} based on calibration training. For calculations-based methods in~\cite{DBLP:conf/infocom/LiuXYWL22}, Liu et al. replaced the process of computing model updates with Quasi-Newton methods to reduce the consumption in the retraining process. Since Federaser~\cite{DBLP:conf/iwqos/LiuMYWL21} can only unlearn a specific client, and given that our scheme specifically targets class-level unlearning, we compare Federaser with our unlearning scheme in the no-IID scenario. For pruning-based (\textbf{Pruning})~\cite{DBLP:conf/www/Wang0XQ22} and calculations-based (\textbf{The Right})~\cite{DBLP:conf/infocom/LiuXYWL22}), they only consider unlearning operations in IID scenario, so we compare those schemes with ours under the IID scenario. Both IID and non-IID scenarios demonstrate the high adaptability of our scheme.

\begin{figure}[!t]
    \centering
    \subfloat[ResNet20]{\includegraphics[width=0.25\textwidth]{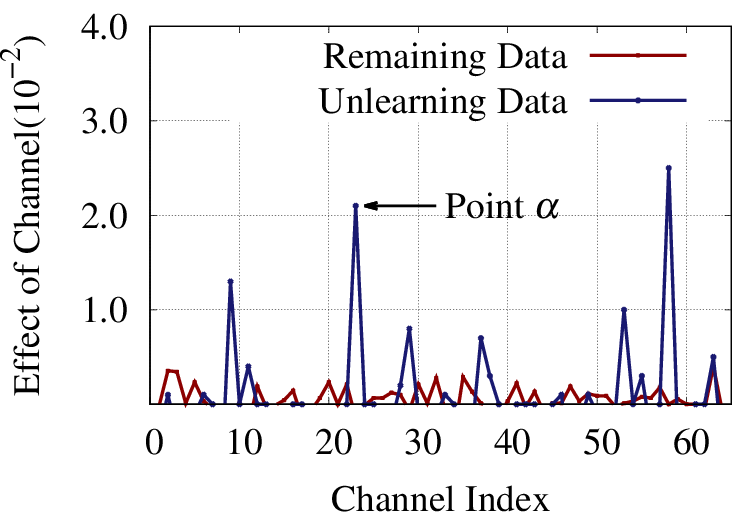}%
    \label{fig:accuracyforpointsresnet20}}
    \subfloat[VGG11]{\includegraphics[width=0.25\textwidth]{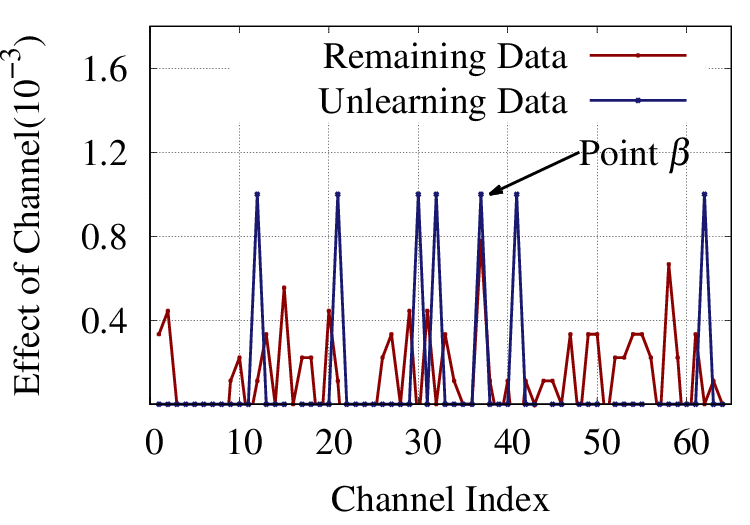}%
    \label{fig:accuracyforpointsvgg11}}
    \caption{The Effect of Each Channel.}
    \label{fig:accuracyforpoints}
\end{figure}

\subsection{Explanation Analysis}
To demonstrate the validity of our scheme, we need to verify whether manipulating a group of parameters in important channels will cause a significant decrease in the performance of the unlearning data. This result indicates that those selected important channels indeed contain information about the unlearning data. Reasonable manipulation that removes this information from the model will achieve the unlearning purpose.

To simulate the real environment settings of federated learning, we evenly distribute the CIFAR-10 datasets to all clients, and the data samples across all clients are IID. We respectively train two models, VGG11 and ResNet20, with the number of participant clients as 10, global epoch = 50, local epoch =  5, local batch size = 128 and learning rate = 0.1. We select the unlearning class as class $1$ and consider other class data to be the remaining data. After the global training process, we first calculate the effect of channels within a layer based on Definition~\ref{def:effectofindividual}. Figure~\ref{fig:accuracyforpoints} presents the effect of each channel with respect to unlearning data and remaining data.

In Figure~\ref{fig:accuracyforpoints}, Figure~\ref{fig:accuracyforpointsresnet20} illustrates the effect of each channel in layer $layer3.block\_2.conv\_bn2.conv$ in model ResNet20, while Figure~\ref{fig:accuracyforpointsvgg11} only illustrates the effect of the first 64 channels in layer $feature.conv\_8$ in model VGG11, since there are totally 512 channels in this layer. It can be seen that when we prune the parameters in one channel at a time, the model performance changes differently for the remaining data and unlearning data. First, when we prune some specific channels, the accuracy of the unlearning data will change dramatically, while the accuracy of the remaining data almost remains stable, such as the channel $23$ in layer $layer3.block\_2.conv\_bn2.conv$~(see the masked point $\alpha$ in Figure~\ref{fig:accuracyforpointsresnet20}). This indicates that some channels, indeed, contain some information learned from the unlearning data; erasing this information will dramatically affect the performance of unlearning data. Second, some channels are prone to affect both unlearning data and remaining data. When we prune those types of parameters, the accuracy of unlearning data and remaining data will simultaneously decrease, such as the channel $37$ in layer $feature.conv\_8$~(see the masked point $\beta$ in Figure~\ref{fig:accuracyforpointsvgg11}). This suggest that some channels affect the performance of both data.

To further illustrate the effect of each channel, we also evaluate how the channel cooperatively affects the unlearning data. To do this, we first select a group of channels based on different ways, then prune the selected channels' parameters and compare the accuracy of the unlearning data and the remaining data. The selection methods include random selection, important selection, and non-important selection. The important selection will choose those channels with higher effect values for unlearning data. The proportion of selection is $\delta = 0.3$. Table~\ref{tab:accuracyforlayer} shows the results of our experiment.

\begin{figure*}[!t]
\centering
    \subfloat[\centering ResNet20 with CIFAR10]{\includegraphics[width=0.25\textwidth]{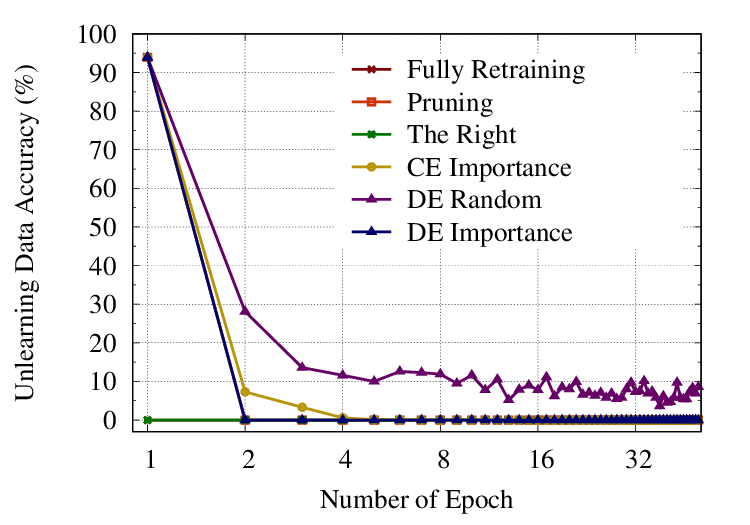}%
    \label{fig:cifar10resnet20unlearning}}
    \subfloat[\centering VGG11 with CIFAR10]{\includegraphics[width=0.25\textwidth]{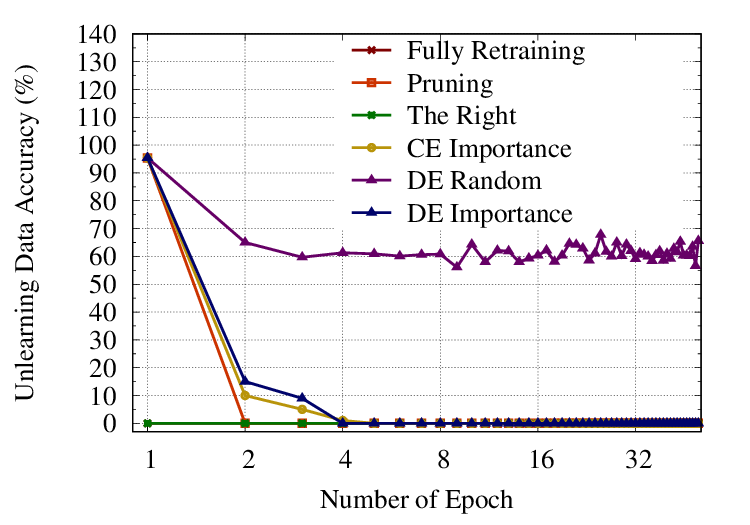}%
    \label{fig:cifar10vgg11unlearning}}
    \subfloat[\centering ResNet20 with CIFAR100]{\includegraphics[width=0.25\textwidth]{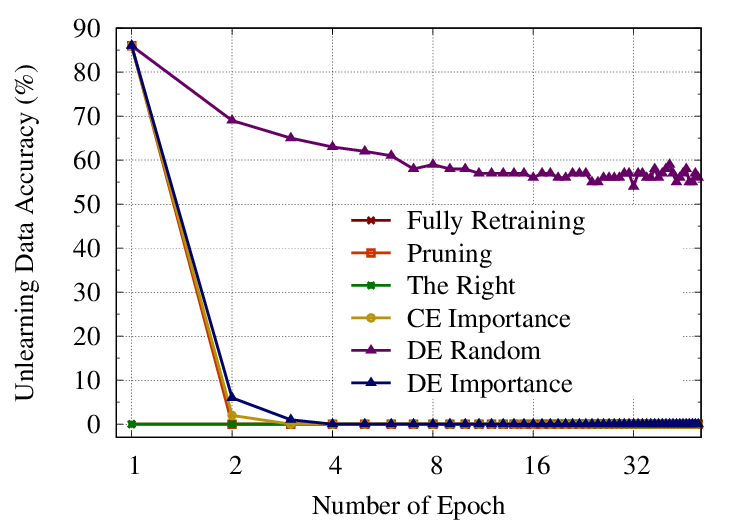}%
    \label{fig:cifar100resnet20unlearning}}
    \subfloat[\centering VGG11 with CIFAR100]{\includegraphics[width=0.25\textwidth]{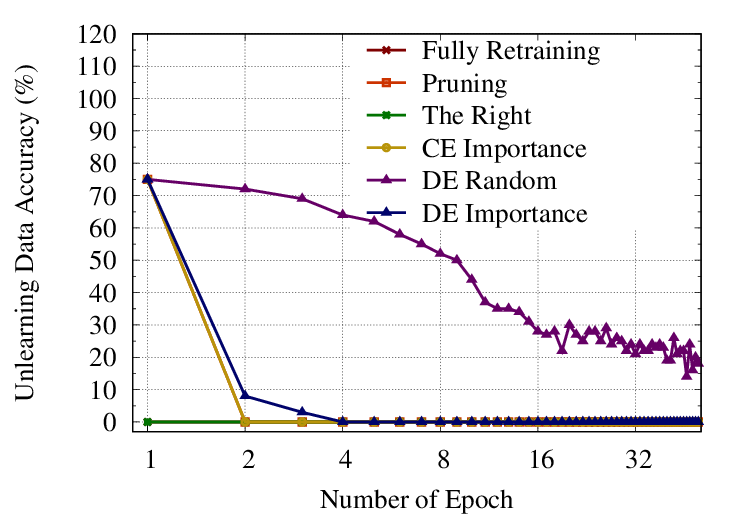}%
    \label{fig:cifar100vgg11unlearning}}\\
\caption{Accuracy of Unlearning Data for Different Unlearning Settings in IID Scenario.}
\label{fig:accuracyunlearningdata}
\end{figure*}

\begin{figure*}[!t]
\centering
    \subfloat[\centering ResNet20 with CIFAR10]{\includegraphics[width=0.25\textwidth]{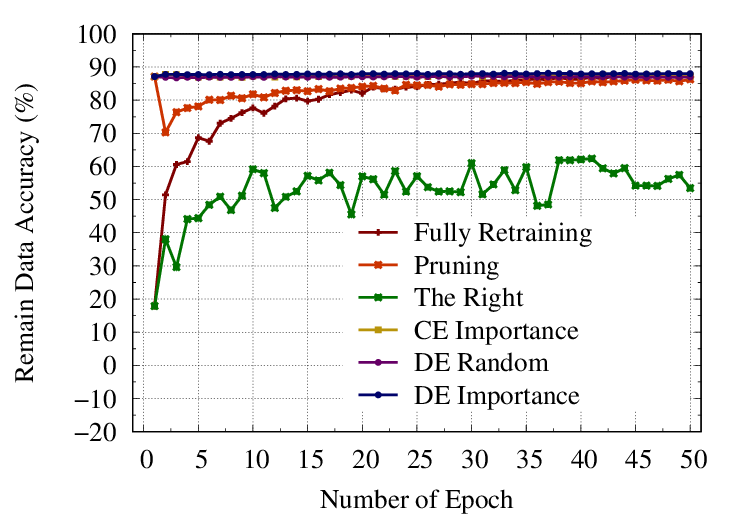}%
    \label{fig:cifar10resnet20remain}}
    \subfloat[\centering VGG11 with CIFAR10]{\includegraphics[width=0.25\textwidth]{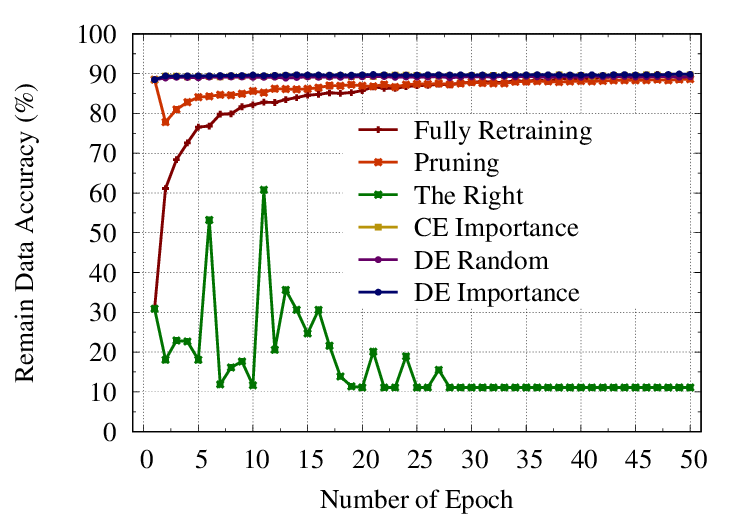}%
    \label{fig:cifar10vgg11remain}}    
    \subfloat[\centering ResNet20 with CIFAR100]{\includegraphics[width=0.25\textwidth]{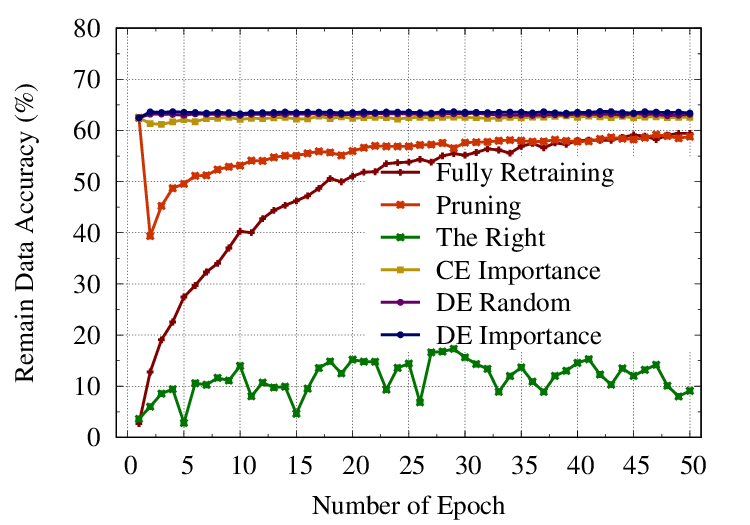}%
    \label{fig:cifar100resnet20remain}}
    \subfloat[\centering VGG11 with CIFAR100]{\includegraphics[width=0.25\textwidth]{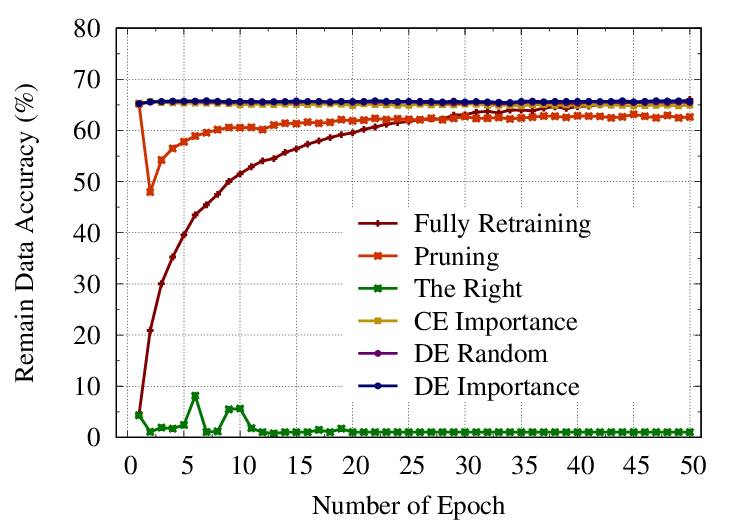}%
    \label{fig:cifar100vgg11remain}}\\
\caption{Accuracy of Remaining Data for Different Unlearning Settings in IID Scenario.}
\label{fig:accuracyremainingdata}
\end{figure*}

\begin{figure*}[!t]
\centering
    \subfloat[Remaining Data(MNIST)]{\includegraphics[width=0.25\textwidth]{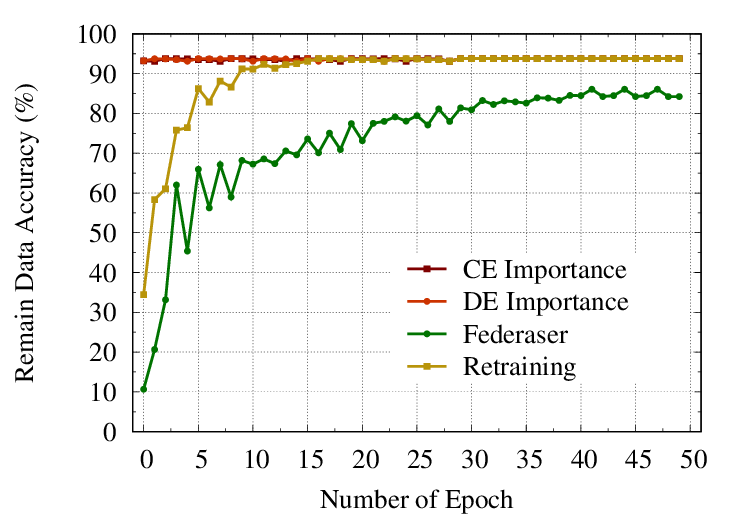}%
    \label{fig:noidd_1}}
    \subfloat[Unlearning Data(MNIST)]{\includegraphics[width=0.25\textwidth]{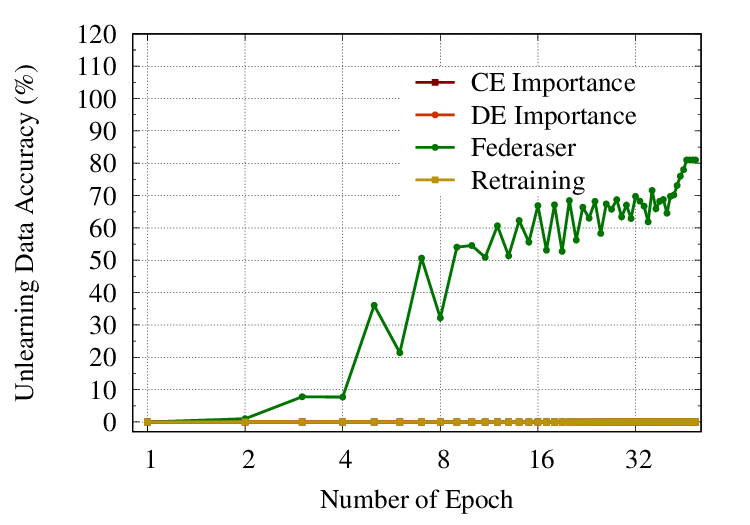}%
    \label{fig:noidd_2}}
    \subfloat[Remaining Data(FMNIST)]{\includegraphics[width=0.25\textwidth]{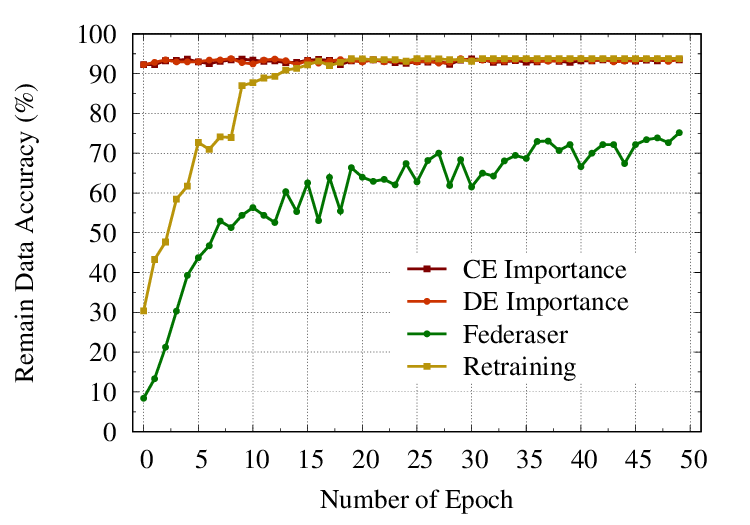}%
    \label{fig:noidd_3}}
    \subfloat[Unlearning Data(FMNIST)]{\includegraphics[width=0.25\textwidth]{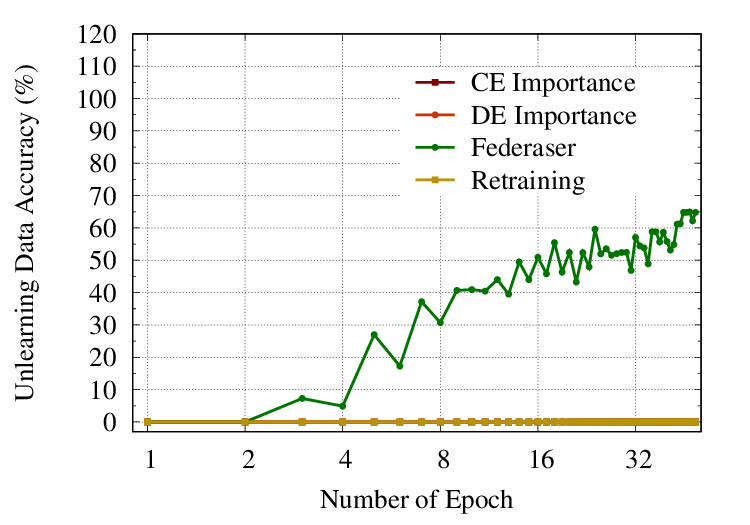}%
    \label{fig:noidd_4}}
\caption{Accuracy of Unlearning and Remaining Data in no-IID Scenario.}
\label{fig:noidd}
\end{figure*}

\begin{table}[]
    \centering
    \renewcommand{\arraystretch}{1.3}
    \caption{Experimental Performance Comparison Between Our Scheme and Retraining From Scratch.}
    \begin{tabular}{ccccc}
    \hline
        &   & Comp (s) & Comm (MB) & Storage\\ \hline
    \multirow{3}{*}{\begin{tabular}[c]{@{}c@{}}ResNet20 \\with\\ CIFAR10\end{tabular}} & Retraining & 5670.39   & 770   & 45000\\
        & DE  & 276.67 & 1.1 & 45000\\
        & CE  & 15.22  & 0   & 1600\\ \hline
    \multirow{3}{*}{\begin{tabular}[c]{@{}c@{}}VGG11 \\with \\CIFAR10\end{tabular}}  & Retraining & 4620.23    & 22800   & 45000\\
        & DE  & 810.78 & 0.1146  & 45000\\
        & CE  & 3.24   & 0   & 1600\\ \hline
    \multirow{3}{*}{\begin{tabular}[c]{@{}c@{}}ResNet20 \\with\\ CIFAR100\end{tabular}}  & Retraining & 8270.05    & 1100    & 49500\\
        & DE  & 843.28 & 3.3 & 49500\\
        & CE  & 5.35   & 0   & 1600\\ \hline
    \multirow{3}{*}{\begin{tabular}[c]{@{}c@{}}VGG11 \\with\\CIFAR10\end{tabular}}  & Retraining & 7480.52    & 30560   & 49500\\
        & DE  & 798.12 & 2.292   & 49500\\
        & CE  & 1.23   & 0   & 1600\\ \hline
    \end{tabular}
    \label{tab:experientialcost}
\end{table}
In Table~\ref{tab:accuracyforlayer}, the layer $layer3.block\_2.conv\_bn2.conv$ and layer $layer3.block\_2.conv\_bn1.conv$ are from ResNet20, while layer $feature.conv\_7$ is from the VGG11. As shown in this table, similar results for the change of unlearning data and remaining data can be found. First, when we prune a group of important parameters, the model performance for the unlearning data will decrease dramatically while the accuracy of the remaining data stays stable. For example, in layer $layer3.block\_2.conv\_bn2.conv$ in ResNet20, the accuracy of unlearning data drops from $87.10\%$ to $38.50\%$, while the remaining data's accuracy only decreases by $80.48\%-70.62\% = 9.86\%$. Second, for random selection or selection of non-important methods, the accuracy of the unlearning and remaining data will decrease simultaneously, and the degree of decrease of the unlearning data is negligible compared to the important selection method. For example, in $feature.conv\_7$ in VGG11, when we prune the randomly selected parameters, the accuracy of the unlearning data decreases by $94.60\%-79.9\% = 14.70\%$, while pruning the important parameters results in $94.60\%-7.30\% = 87.30\%$. In addition, if we prune the important parameters in the intermediate layer of one model, the accuracy of the remaining data also drops a lot. For example, in layer $layer3.block\_2.conv\_bn1.conv$ in ResNet20, pruning the important parameters causes the accuracy of the remaining data to drop by $80.48\%-65.14\% = 15.34\%$. This illustrates that, in some model layers, especially those near the front part of the model, parameters that are important for the unlearning data may also be important for the remaining data. For example, the important parameters for unlearning data used to detect textures or materials may also be used to detect the same textures or materials in the remaining data in a CNN layer. If we prune the important parameters directly to achieve the unlearning purpose, that may impact the accuracy of the remaining data~\cite{DBLP:conf/www/Wang0XQ22}. 

\textbf{Summary:} Based on the above experimental results, the most important parameters of the model for unlearning data can be identified based on the ablation study. Reasonable manipulation of these parameters will achieve the purpose of unlearning data. Also, we find that those parameters that affect the unlearning data may also have a large impact on the remaining data. Therefore, in our unlearning scheme, to ensure the performance of the unlearned model, we maintain all the parameters of the global model and only fine-tune important parameters to unlearn data rather than directly prune all important parameters.

\subsection{Performance Analysis}
\label{sec:performanceanalysis}

\subsubsection{Accuracy Evaluation}
\label{sec:accuracyevaluation}

\begin{figure}[!t]
    \centering
    \subfloat[Original Model]{\includegraphics[width=0.23\textwidth]{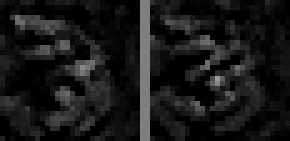}}
    \subfloat[Fully Retraining]{\includegraphics[width=0.23\textwidth]{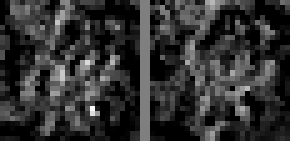}}\\
    \subfloat[Decentralized Unlearning]{\includegraphics[width=0.23\textwidth]{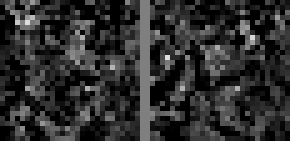}}
    \subfloat[Centralized Unlearning]{\includegraphics[width=0.23\textwidth]{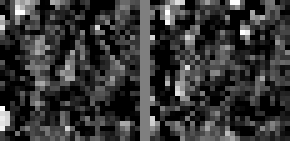}}
    \caption{The Results of Model Inversion Attack}
\label{fig:unlearningdataaccuracy}
\end{figure}

To demonstrate the effectiveness and efficiency of our unlearning scheme, we measure the accuracy of the unlearning data and the remaining data during the unlearning process, respectively. We first train two models, ResNet20 and VGG11, respectively, using two datasets, CIFAR-10 and CIFAR-100. The configuration of the training process is as follows: client number = 10, global epoch = 50, local epoch =  5, local batch size = 128 and learning rate = 0.1. The data samples across all clients are IID. After the global training process, we unlearn the class $1$ from two models respectively, using our unlearning scheme in three different settings, CE Importance, DE Random and DE Importance. For CE Importance, we put 1600 samples on the server side to simulate the training data owned by the server and set learning rate = 0.01 and batch size = 64 to fine-tune models. For $\delta$, we set $\delta = 0.05$ for ResNet20, and $\delta = 0.01$ for VGG11. The other setting for the fine-tuning process is the same as the original global training process. We compare the results with three different methods: fully retraining, pruning-based methods~\cite{DBLP:conf/www/Wang0XQ22} and the right~\cite{DBLP:conf/infocom/LiuXYWL22}. For the pruning-based and the right method, we set all hyperparameter settings suggested in their paper. For fully retraining, we set hyperparameters to be the same as the original training process. 

\begin{table}[!t]
\caption{The Results of the MIAs}
\label{tab:membershipattackresults}
\centering
    \renewcommand{\arraystretch}{1.3}
    \begin{tabular}{ccc}
    \hline
    \rowcolor{gray!20}                  & CIFAR-10 + ResNet    & MNIST + ResNet  \\ \hline
    Original                            & 95.62\%               & 94.56\%               \\
    \rowcolor{gray!20}Fully Retraining  & 0.00\%                & 0.00\%                \\
    CE important                        & 0.00\%                & 0.00\%                \\
    \rowcolor{gray!20}DE important      & 0.00\%                & 0.00\%                \\ \hline
    \end{tabular}
\end{table}

The results are shown in Figure~\ref{fig:accuracyunlearningdata} and Figure~\ref{fig:accuracyremainingdata}, for fully retraining and the right~\cite{DBLP:conf/infocom/LiuXYWL22}, the x-axis represents the number of retraining epochs; for the pruning-based method~\cite{DBLP:conf/www/Wang0XQ22}, CE Importance, DE Random and DE Importance, the x-axis represents the number of training epochs in the fine-tuning process. We also scale the x-axis to $log_2^{x}$ for convenience in Figure~\ref{fig:accuracyunlearningdata}. The y-axis in Figure~\ref{fig:accuracyunlearningdata} represents the accuracy of unlearning data (class $1$), while in Figure~\ref{fig:accuracyremainingdata}, it represents the accuracy of the remaining data. For the data to be unlearned, CE Importance and DE Importance basically reduce the accuracy to $0$ within $4$ epochs, which means that the model cannot accurately identify the class $1$ after almost $4$ epochs of the unlearning process. For DE Random, it is difficult to reduce the accuracy to $0$ since the fine-tuning parameters are randomly selected, not those with the most influence. For the pruning-based method~\cite{DBLP:conf/www/Wang0XQ22}, since it prunes all the parameters about the unlearning class $1$, it has substantially low accuracy in the beginning. The fully retraining and the right~\cite{DBLP:conf/infocom/LiuXYWL22} method also has a precision of $0$ for the unlearning class since its training set does not contain unlearning data.

However, the other three models, fully retraining, pruning-based~\cite{DBLP:conf/www/Wang0XQ22} and the right~\cite{DBLP:conf/infocom/LiuXYWL22}, do not meet the requirements of unlearning for the accuracy of the remaining data. For the accuracy of the remaining dataset in Figure~\ref{fig:accuracyremainingdata}, the accuracy of CE Importance and DE Importance does not degrade significantly during the unlearning process. This is because we fine-tune only part of the channels while keeping all parameters of the previous model, particularly parameters unrelated to the unlearned class $1$. For fully retraining, since it trains the model from the beginning, the model is only available when the training process is completely finished. For the pruning-based method~\cite{DBLP:conf/www/Wang0XQ22}, in their unlearning scheme, while the parameters associated with class $1$ are pruned, the parameters with other classes also become incomplete, and their accuracy will decrease. Therefore, the unlearned model is only available when the fined-tuned training process is complete. Note that the model accuracy of the pruning-based method~\cite{DBLP:conf/www/Wang0XQ22} is generally higher than that of fully retraining since some of the parameters of the previous model (which are not pruned) are retained in~\cite{DBLP:conf/www/Wang0XQ22}. The right~\cite{DBLP:conf/infocom/LiuXYWL22} is also based on retraining methods, but unlike fully retraining, in the retraining process, they frequently used Quasi-Newton to calculate the update of the global model instead of using the training method. Calculated-based methods are usually applicable to simple models. When faced with complex models with a huge amount of training data, it is difficult for the model to reach an ideal state.

In addition, we also evaluate the effectiveness of our unlearning scheme in no-IID scenario. We choose two datasets, MNIST and Fashion MNIST, for this experiment and assign one class of data per user. The configuration of the training process is as follows: client number = 10, global epoch = 20, local epoch = 5, local batch size = 64 and learning rate = 0.1. For CE Importance, we put 1600 samples on the server side to simulate the training data owned by the server and set the learning rate = 0.1 and batch size = 64 to fine-tune models. For $\delta$, we set $\delta = 0.2$ for both datasets. We also compare the results with Federaser~\cite{DBLP:conf/iwqos/LiuMYWL21} and set all hyperparameter settings suggested in their paper to construct our experiments. The results are shown in Figure~\ref{fig:noidd}.

\begin{figure}[!t]
    \centering
    \subfloat[Original Image]{\includegraphics[width=0.48\textwidth]{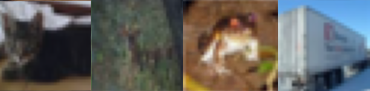}}\\
    \subfloat[Image Generated in Fully Retraining]{\includegraphics[width=0.48\textwidth]{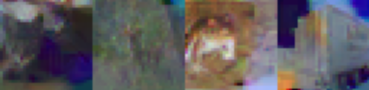}}\\
    \subfloat[Image Generated in Decentralized Unlearning]{\includegraphics[width=0.48\textwidth]{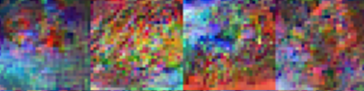}}\\
    \caption{The Results of GradAttack}
\label{fig:gradattack}
\end{figure}

From Figure~\ref{fig:noidd}, we can see that our scheme can also effectively unlearn class information under no-IID scenario and ensure the performance of the model for the remaining data. As for Federaser~\cite{DBLP:conf/iwqos/LiuMYWL21}, since the model still has information about classes that need to be unlearned in the calibration training, the unlearning results are not good.

As can be seen from all subplots in Figures~\ref{fig:accuracyunlearningdata}, ~\ref{fig:accuracyremainingdata} and Figure~\ref{fig:noidd}, our method can remove the contribution of the unlearning class and only use a smaller epoch while simultaneously ensuring the accuracy of the remaining dataset. More importantly, during each epoch, our scheme only updates part of the model parameters, and the amount of training data used by the server is much smaller than other solutions. To further demonstrate the efficiency of our scheme, we record the cost during the entire unlearning process, and the results are shown in Table~\ref{tab:experientialcost}. From those results, the proposed unlearning schemes are found to be efficient in terms of computational, communication, and storage costs compared to retraining from scratch. For example, in our CE Importance unlearning scheme, each epoch only costs $15.22s$ and only needs $1,600$ images for ResNet20 with the CIFAR10 dataset, while fully retraining needs $45,000$ images and costs $5670.39s$. Overall, our unlearning simultaneously considers both effectiveness and efficiency.

\subsubsection{Model Inversion Attack}
To further analyze the effectiveness of our scheme, we construct the following experiment based on model inversion attack~\cite{DBLP:conf/ccs/FredriksonJR15} and membership inference attack~\cite{DBLP:conf/csfw/YeomGFJ18}.  We implemented the model inversion attack as described in~\cite{DBLP:conf/ccs/FredriksonJR15}. We first train model ResNet18 based on the MNIST dataset with the number of participant clients as 10, global epoch = 10, local epoch = 5, local batch size = 128 and learning rate = 0.1. We select the unlearning class as class 3 and consider other class data as the remaining data. After the global training process, we select the most influential parameters for unlearning class $3$ and set the selection factor $\delta = 0.2$. For decentralized unlearning and the fully retraining schemes, we set the hyper-parameters the same as the original training process. For centralized unlearning, we set the local training epoch = 5, lr = 0.005. Figure~\ref{fig:unlearningdataaccuracy} depicts the results of model inversion attacks against models affected by different unlearning methods.

As shown in Figure~\ref{fig:unlearningdataaccuracy}, the images generated by fully retraining, our centralized unlearning, and decentralized unlearning schemes are dark and jumbled since the model inversion attack has relatively little gradient information to rely on. As expected, fine-tuning the influential channels based on perturbation data makes it impossible to infer any significant information about the unlearning data. This implies that the information of unlearning data is almost entirely removed by our centralized and decentralized unlearning schemes, virtually eliminating the possibility of obtaining valuable class information via model inversion attacks.

\subsubsection{Membership Inference Attack}
We also leverage Membership Inference Attacks (MIAs) in paper~\cite{DBLP:conf/csfw/YeomGFJ18}, to assess whether the unlearning data are still identified in the training dataset. We set the number of shadow models as 20 and the training epoch of the shadow model as 10, batch size = 64. In the context of MIAs,
a shadow model is a surrogate model trained to mimic the behavior of the target model that is being attacked. The attack model is a fully connected network with two hidden linear layers of width 256 and 128, respectively, with ReLU activation functions and a sigmoid output layer. We evaluate our unlearning scheme with two settings, CIAFR-10 + ResNet and MNIST + ResNet. We set the unlearning class to $3$. Other configurations of the training process are the same as those described in Section~\ref{sec:accuracyevaluation}. Table~\ref{tab:membershipattackresults} shows the results of our evaluation. 

As seen from the table, for all original models, MIA has a high success rate; i.e., it was able to derive the training dataset containing unlearning data successfully. However, for all other methods, the success rate of MIAs is lower, which indicates that MIA cannot deduce the existence of unlearning data in the training dataset; this indicates that the effect of the unlearning data has been successfully removed from the unlearned model. 

\textbf{Summary:} The above experiments show that our scheme can effectively maintain the performance of the unlearned model, and reduce the probability of an attacker obtaining confidential information about those targets.

\subsection{Gradient Attack}

In our decentralized unlearning scheme, we only upload the updates of partial channels' parameters. In order to evaluate whether updating partial parameters can alleviate the risks caused by gradients, we use GradAttack~\cite{DBLP:conf/nips/HuangGSLA21} to attack the training process of fully retraining and our decentralized unlearning, respectively. We use the \textit{realistic setting} in their paper to attack ResNet20 with CIAFR-10 and set the unlearning data as class $1$ and $\delta = 0.05$ in our decentralized unlearning setting. Other experimental settings are the same as the experiments in~\ref{sec:accuracyevaluation}. Figure~\ref{fig:gradattack} shows part of the results of our experiments.

As shown in Figure~\ref{fig:gradattack}, for the fully retraining scheme, since it uploads all the parameters' updates, some information of original data can be recovered using the gradients. For our decentralized unlearning, it is impossible to obtain any information from the gradient-based generated images since we only upload partial gradients. \textbf{In summary}, our decentralized unlearning scheme not only reduces the consumption caused by communication but also alleviates the security risks caused by gradients.

\begin{figure}[!t]
\centering
    \subfloat[Unlearning Data]{\includegraphics[width=0.25\textwidth]{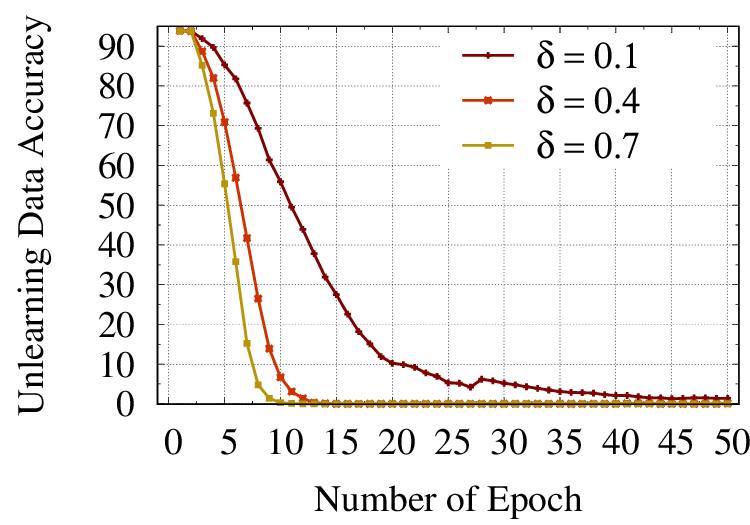}%
    \label{fig:delataeffectaccuracyunlearning}}
    \subfloat[Remaining Data]{\includegraphics[width=0.25\textwidth]{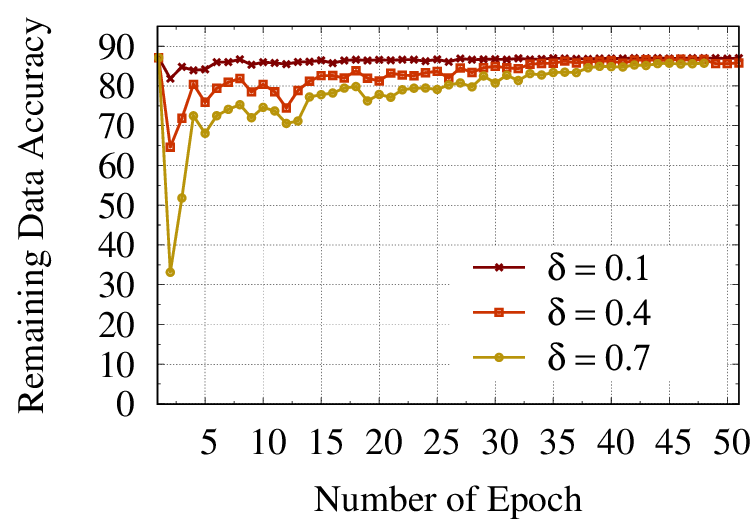}%
    \label{fig:delataeffectaccuracyremaining}}
\caption{The Effect of $\delta$ in Different Settings}
\label{fig:delataeffect}
\end{figure}

\subsection{The Effect of Selection Factor $\delta$}
The hyperparameter $\delta$ determines how many influential channels are selected. To evaluate the effect of $\delta$, we use the above experimental setting, in which the dataset is CIFAR-10, and the model is ResNet20, to train the global model. After that, we execute a centralized unlearning process with different $\delta$. Figure~\ref{fig:delataeffectaccuracyunlearning} and Figure~\ref{fig:delataeffectaccuracyremaining} show the results on the unlearning data and remaining data, respectively. It can be seen from Figure~\ref{fig:delataeffectaccuracyunlearning}, that selecting more influential channels can accelerate the unlearning process due to the increasing number of pruned and fine-tuned channels. However, as depicted in Figure~\ref{fig:delataeffectaccuracyremaining}, increasing the number of pruned channels will decrease the performance of the model after the pruning process. This will result in the need for more fine-tuning to recover model performance. Meanwhile, more channels mean the consumption of updates to uploads will increase in the decentralized unlearning process. Therefore, in principle, the setting of the selection factors should consider the usability of the unlearned model and the efficiency of the unlearning process.

\section{Conclusion and future work}
\label{sec:conclusion}
In this paper, we proposed a federated unlearning scheme that addresses the practical problem of removing the effects of particular classes from a trained model in the federated learning context.  As a solution, we analyzed the most influential channels of a model for the given classes based on the ablation study. For unlearning class, we provided two effective methods that only fine-tune partial influential channel parameters with the help of perturbation data. The analysis and experimental results demonstrate that under our scheme, the model can remove the impact of classes that need to be removed and ensure the accuracy of the remaining data in a quick and efficient manner. 

In future work, we plan to explore more complex federated learning scenarios containing more kinds of unlearning requests, for example, unlearning one sample or all data in one client. At present, we have only illustrated that there are different influences within a trained model's channel at the class level. Due to the various unlearning requests involved, we intend to explore ways to extend and/or modify the current method of evaluating channel importance and develop new federated unlearning schemes based on these revised evaluations. In addition, we plan to design a more powerful scheme that supports unlearning requests from Natural Language Processing~(NLP) models, combined with other technologies such as information theory.

\section{Acknowledgement}
This work was supported in part by NSF under grants III-2106758 and POSE-2346158.

\bibliographystyle{IEEEtran}
\bibliography{bare_jrnl_new_sample4}

\vspace{-4em}
\begin{IEEEbiography}[{\includegraphics[width=1in,height=1.25in,clip,keepaspectratio]{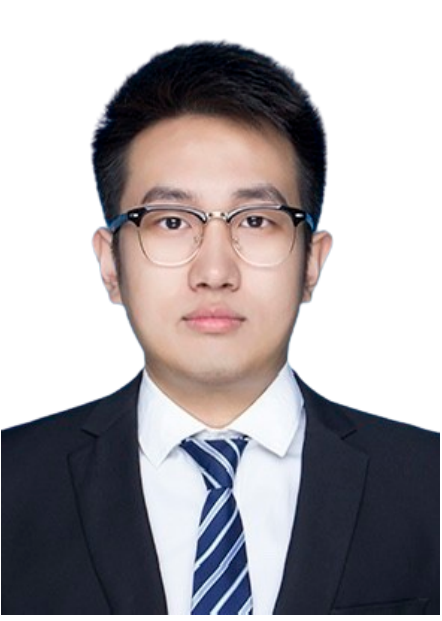}}]{Heng Xu} received his B.Eng. and M.Eng. degree from China University of Geosciences, China, in 2018 and 2021, respectively. He is currently a PhD student in University of Technology Sydney, Australia. His research interests are machine unlearning.
\end{IEEEbiography}

\vspace{-2em}
\begin{IEEEbiography}[{\includegraphics[width=1in,height=1.25in,clip,keepaspectratio]{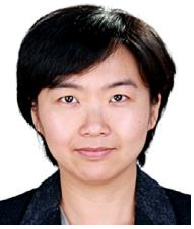}}]{Tianqing Zhu}received her BEng and MEng degrees from Wuhan University, China, in 2000 and 2004, respectively, and a PhD degree from Deakin University in Computer Science, Australia, in 2014. Dr. Tianqing Zhu is currently a professor in the faculty of data science at the City University of Macau. Before that, she was a lecturer at the School of Information Technology, Deakin University, Australia, and an associate professor at the University of Technology Sydney, Australia. Her research interests include privacy-preserving and AI security. 
\end{IEEEbiography}

\begin{IEEEbiography}[{\includegraphics[width=1in,height=1.25in,clip,keepaspectratio]{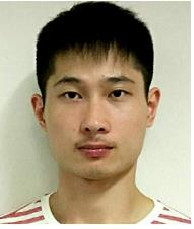}}]{Lefeng Zhang} received his B.Eng. and M.Eng. degree from Zhongnan University of Economics and Law, China, in 2016 and 2019, respectively; and a PhD degree from University Technology Sydney, Australia, in 2024. He is currently an assistant professor at City University of Macau. His research interests are game theory and privacy-preserving.
\end{IEEEbiography}

\begin{IEEEbiography}[{\includegraphics[width=1in,height=1.25in,clip,keepaspectratio]{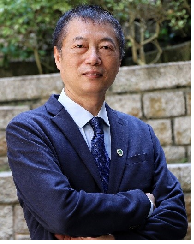}}]{Wanlei Zhou} (Senior member, IEEE) is currently the Vice Rector (Academic Affairs) and Dean of the Faculty of Data Science, City University of Macau, Macao SAR, China. He received the B.Eng and M.Eng degrees from Harbin Institute of Technology, Harbin, China in 1982 and 1984, respectively, and the PhD degree from The Australian National University, Canberra, Australia, in 1991, all in Computer Science and Engineering. He also received a DSc degree (a higher Doctorate degree) from Deakin University in 2002. Before joining City University of Macau, Professor Zhou held various positions including the Head of School of Computer Science in University of Technology Sydney, Australia, the Alfred Deakin Professor, Chair of Information Technology, Associate Dean, and Head of School of Information Technology in Deakin University, Australia. Professor Zhou also served as a lecturer in University of Electronic Science and Technology of China, a system programmer in HP at Massachusetts, USA; a lecturer in Monash University, Melbourne, Australia; and a lecturer in National University of Singapore, Singapore. His main research interests include security, privacy, and distributed computing. Professor Zhou has published more than 400 papers in refereed international journals and refereed international conferences proceedings, including many articles in IEEE transactions and journals. 
\end{IEEEbiography}

\begin{IEEEbiography}[{\includegraphics[width=1in,height=1.25in,clip,keepaspectratio]{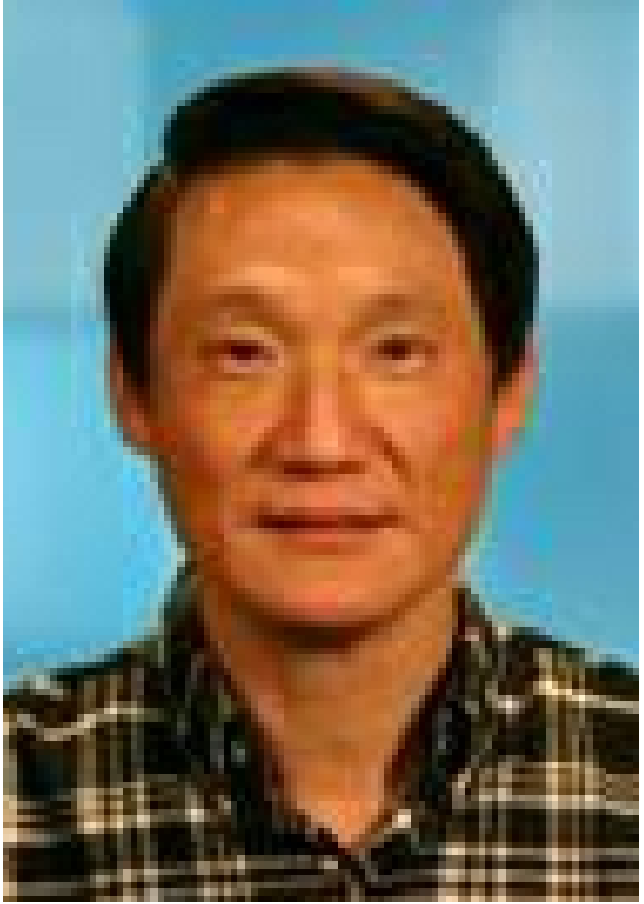}}]{Philip S. Yu} is a Distinguished Professor at the University of Illinois at Chicago, holding the Wexler Chair in Information and Technology. Previously, he managed the Software Tools and Techniques department at IBM Thomas J. Watson Research Center. He is a Fellow of the ACM and IEEE. He received the ACM SIGKDD 2016 Innovation Award, the IEEE Computer Society's 2013 Technical Achievement Award, the Research Contributions Award from IEEE ICDM in 2003, and an IEEE Region 1 Award in 1999. He has also been honored with several UIC and IBM awards, including Research of the Year (2013), UI Faculty Scholar (2014), and numerous IBM accolades, such as Outstanding Innovation Awards and an IBM Master Inventor title. He is the Editor-in-Chief of ACM Transactions on Knowledge Discovery from Data and serves on the steering committee of ACM Conference on Information and Knowledge Management. He was also the Editor-in-Chief of IEEE Transactions on Knowledge and Data Engineering (2001-2004) and has served as an associate editor for multiple journals. He has been actively involved in numerous conferences as a program chair, co-chair, or committee member, and has held leadership roles in major conferences like IEEE Intl. Conf. on BIGDATA (2016), IEEE Intl. Conf. on Data Science and Advanced Analytics (2014), and IEEE/ACM Intl. Conf. on Advances in Social Networks Analysis and Mining (2012). He has published over 970 papers, has more than 197,002 citations with an H-index of 197, and holds or has applied for over 300 US patents. His main research interests include big data, data mining (especially on graph/network mining), social network, privacy preserving data publishing, data stream, database systems, and Internet applications and technologies.
\end{IEEEbiography}

\end{document}